\pdfoutput=1 

\documentclass[aps,pre,showpacs,twocolumn]{revtex4-1}

\usepackage{amsmath,amssymb,amsfonts}
\usepackage{graphicx,epsf}
\usepackage{xspace}
\usepackage{textcomp}
\usepackage{color}
\usepackage[normalem]{ulem}

\usepackage[colorlinks,bookmarks=false,citecolor=blue,linkcolor=blue,urlcolor=blue,hyperfootnotes=true]{hyperref}

\newcommand{\be}{\begin{equation}}
\newcommand{\ee}{\end{equation}}

\newcommand{\p}{\partial} 
\newcommand{\vx}{\vec{x}}

\newcommand{\bq}{{\bf q}}

\newcommand{\bx}{{\bf x}}
\newcommand{\vq}{{\vec{q}}}
\newcommand{\vp}{{\vec{p}}}
\newcommand{\td}{\text{\tiny $D$}}
\newcommand{\xx}{\text{\tiny $X$}}

\newcommand{\xn}{\text{\tiny $0$}}

\newcommand{\lr}{\text{\tiny $\mathrm{LR}$}}
\newcommand{\sr}{\text{\tiny $\mathrm{SR}$}}
\newcommand{\gt}{\text{\tiny $\mathrm{T}$}}

\newcommand{\ew}{\text{\tiny $\mathrm{EW}$}}
\newcommand{\ewlr}{\text{\tiny $\mathrm{EWLR}$}}

\newcommand{\tih}{\tilde{h}}
\newcommand{\tj}{\tilde{j}}

\newcommand{\tp}{\hat{p}}
\newcommand{\tq}{\hat{q}}

\newcommand{\tg}{\hat{g}}
\newcommand{\tf}{\hat{f}}
\newcommand{\tI}{\hat{I}}

\newcommand{\etan}{\eta^{\nu}}
\newcommand{\etad}{\eta^{\td}}
\newcommand{\etax}{\eta^{\xx}}

\newcommand{\dis}{\displaystyle}

\newcommand{\anz}{{ansatz}\xspace}

\newcommand{\tm}{\text{-}}

\begin{document}

\title{The Kardar-Parisi-Zhang equation with spatially correlated noise: a unified picture from nonperturbative renormalization group}

\author{Thomas Kloss$^1$, L\'eonie Canet$^2$, Bertrand Delamotte$^3$, and  Nicol\'as Wschebor$^{3,4}$}
\affiliation{
$^1$International Institute of Physics, UFRN, Av.\ Odilon Gomes de Lima 1722, 59078-400 Natal, Brazil \\
$^2$  LPMMC, CNRS UMR 5493, Universit\'e Joseph Fourier Grenoble, Bo\^ite Postale 166,  38042 Grenoble, France\\
$^3$LPTMC, CNRS UMR 7600, Universit\'e Pierre et Marie Curie, Bo\^{i}te Postale 121, 75252 Paris, France\\
$^4$Instituto de F\'isica, Facultad de Ingenier\'ia, Universidad de la Rep\'ublica, J.H.y Reissig 565, 11000 Montevideo, Uruguay}

\begin{abstract}

We investigate the scaling regimes of the  Kardar-Parisi-Zhang equation
 in the presence of  spatially correlated noise with power law decay  $D(p) \sim
p^{-2\rho}$ in Fourier space, using a nonperturbative renormalization
group  approach. We determine the full phase diagram of the system as
a function of $\rho$ and the dimension $d$.  In addition to the
weak-coupling part of the diagram, which agrees with the results 
from Refs.\  [Europhys.\ Lett.\ {\bf 47}, 14 (1999), Eur.\ Phys.\ J.\ B
  {\bf 9}, 491 (1999)], we find the two fixed points describing the
short-range (SR) and long-range (LR) dominated strong-coupling
  phases. In contrast with a suggestion in the references cited above, we
  show that, for all values of $\rho$, there exists a unique strong-coupling SR fixed point  that can be continuously followed as a function of $d$. We show 
in particular that the existence and the behavior of the LR fixed point do not provide 
any hint for 4 being the upper critical dimension of the KPZ equation with SR noise.
\end{abstract}
\maketitle

\section{Introduction}

To describe interface roughening and its dynamical scaling,  Kardar, Parisi and Zhang (KPZ) proposed  a nonlinear Langevin equation, 
which has now emerged as a fundamental model to study nonequilibrium phase transitions and scaling phenomena \cite{kardar86,halpin-healy95,krug97,Barabasi95}.
The KPZ equation \cite{kardar86} writes 
\begin{equation}
\frac{\p h(t,\vx)}{\p t} = \nu\,\nabla^2 h(t,\vx) \, + 
\,\frac{\lambda}{2}\,\big(\nabla h(t,\vx)\big)^2 \,+\,\eta(t,\vx) ,
\label{eqkpz}
\end{equation}
where $h(t,\vx)$ is a single valued height profile which depends on the $d$-dimensional spatial coordinate $\vx$ of 
the substrate and on time $t$, $\nu$ the surface tension, and 
 $\eta(t,\vx)$ represents a Gaussian noise with zero mean $\langle \eta(t,\vx)\rangle = 0$ and variance
\begin{equation}
  \big\langle \eta(t,\vx)\eta(t',\vx')\big\rangle = 2\, {\cal D}(\vx-\vx')\,\delta(t-t'). 
\end{equation}
 The non-linear term proportional to  $\lambda$ is the essential ingredient to capture the  
 dynamical roughening of the interface \cite{halpin-healy95,krug97,Barabasi95}. 

The original KPZ equation is formulated with a purely  local noise of amplitude $D$, that is ${\cal D}(\vx-\vx')= D\delta^d(\vx-\vx')$.
 This equation encompasses the following behavior. It always  generates scaling in the stationary regime, characterized by the dynamical  
$z$ and the roughness $\chi$ critical exponents. For dimensions $d\leq 2$, the interface always roughens,
 whereas for $d>2$, a nonequilibrium phase transition occurs for a critical value  $\lambda_c$ of the non-linearity, which 
 separates a strong-coupling ($\lambda>\lambda_c$) rough phase from a weak-coupling  ($\lambda <\lambda_c$) smooth phase 
 corresponding to the linear Edwards Wilkinson (EW) regime, with exponents $z=2$ and $\chi=(d-2)/2$. The ubiquity of the 
 KPZ universality class  has led to considerable efforts  over the last decades to understand its statistical 
 properties \cite{halpin-healy95,krug97}. We do not review here all the corresponding literature, but only mention 
 the most recent contributions. For one-dimensional interfaces, an impressive breakthrough has been achieved during 
 the last years both theoretically \cite{Praehofer04,Sasamoto05,Sasamoto10,Calabrese11,Amir11,Imamura12} (and for a 
 review, see {\it e.g.} \cite{Corwin12}) and experimentally \cite{Takeuchi10,Takeuchi11,Takeuchi12}. 
 For higher-dimensional interfaces, recent large scale numerical simulations were launched to refine the    
 estimates of critical exponents and  probability distributions \cite{Kelling11,Pagnani13,Halpin-Healy12,Halpin-Healy13,*Halpin-Healy13err,Oliveira13}. 
However, the progress is much slower, leaving still unsettled debates such as the existence of an upper critical dimension for this model. 

 Recently, we proposed a nonperturbative renormalization group (NPRG) approach for the KPZ equation
 \cite{Canet10,Canet11a,*Canet12Err,Kloss12}, which successfully yielded the fully attractive (short-range (SR)) strong coupling 
fixed point describing the rough phase in all dimensions. The associated exponents are in close (resp.\ reasonable) agreement
in $d=2$ (resp.\ $d=3$) with the estimates from numerical simulations \cite{Tang92b,Ala-Nissila93,Castellano99,Marinari00,Reis04,Ghaisas06,Kelling11,Pagnani13}. The finding of the fully attractive strong-coupling fixed-point allows one to show the emergence of generic scaling for the 2-point correlation  and response functions.   The resulting scaling functions in $d=1$ compare remarkably well with the exact results \cite{Canet11a,*Canet12Err,Praehofer04}.  These calculations have been extended in any dimensions,
 giving in particular the 2-point correlation and response functions  in $d=2$ and $d=3$ \cite{Kloss12}.
 The ensuing predictions for the associated universal amplitude ratios in $d=2$ have been recently accurately confirmed in lattice simulations \cite{Halpin-Healy13}. 

We here address the issue of the presence of long-range (LR) correlated noise in the KPZ equation. Some experimental realizations (such as wetting in porous media \cite{Rubio89,Horvath91}) suggested that spatial correlations may exist at the microscopic level, in
the noise or in the hydrodynamical interactions \cite{halpin-healy95,krug97,Barabasi95}. This has triggered the study of the relevance of this type of microscopic correlations,  as for its impact on the critical exponents and  on the phase diagram.
  Several numerical and theoretical studies have shown that  spatial 
\cite{Meakin89,Hentschel91,Amar91,Peng91,Pang95,Li97,Mukherji97,Janssen99,Frey99,Katzav99,Verma00,Katzav03,Katzav13} and/or temporal \cite{Medina89,Lam92,Katzav04,Fedorenko08} 
noise correlations indeed lead to new phases with modified exponents. 
Following Ref.\ \cite{Medina89,Janssen99,Frey99,Katzav99}, we consider, in addition to the local delta-correlated SR noise, a spatially correlated noise of the form
\begin{equation}
  {\cal D}_{\lr}(\vx-\vx') \sim |\vx-\vx'|^{2 \rho - d}  , \quad   \rho \leq d / 2.
  \label{eq:lr-noise}
\end{equation}
More precisely, the full noise term writes in Fourier space
\begin{equation}
 {\cal D}(\vp) = D(1+w p^{-2 \rho}). 
  \label{eq:lr-noise-fourier1}
\end{equation}
where $p=|\vp|$ and $w$ is the relative amplitude of the LR noise.

The early dynamical renormalization group (DRG) analysis by Medina {\it et al.} \cite{Medina89} predicted the existence of a rough LR dominated phase above a threshold value of the decay exponent $\rho$ of the LR noise,  with associated $\rho$-dependent critical exponents $\chi = (2-d+2\rho)/3$ and $z=2-\chi$. This prediction
  was confirmed by a functional RG calulcation for directed polymers \cite{Halpin-Healy90}. Yet some other theoretical approaches, based on a replica scaling analysis \cite{Zhang90} or on a scaling analysis in open dissipative  systems \cite{Hentschel91} yielded alternative predictions for the critical exponents and the threshold value of $\rho$. As early numerical simulations, mainly in one dimension, were not in accordance, the situation was unclear. However, later simulations \cite{Hayot96} of the Burgers equation in $d=1$ clearly confirmed the original DRG results, which were then also supported by
 a Mode-Coupling calculation \cite{Bhattacharjee98}, a Self-Consistent Expansion \cite{Katzav99,Katzav13} (at least in $d=1$), and exact results from  a DRG calculation using a stochastic Cole-Hopf transformation by Janssen, Frey, and T\"auber (JFT) \cite{Janssen99,Frey99}. We present below the findings of JFT, which will serve as a reference for later comparison with our work.

JFT have shown that, in the presence of LR noise, new 
LR dominated weak-coupling phases exist. They also suggested the existence of a LR dominated strong coupling phase even if the perturbative analysis cannot  find the associated fixed point.
Furthermore, they derived exact ({\it i.e}.\ valid to all orders in perturbation theory) expressions 
for the corresponding $\rho$-dependent exponents, including the LR dominated strong-coupling phase (under the assumption that 
the associated  fixed-point exists), which coincide with the DRG one-loop result. The physical picture emerging from their work is as follows. Below a lower critical dimension 
$d_c(\rho) = 2(1+\rho)$, no smooth phase is stable, that is, the interface is always rough and the LR noise
is either irrelevant at moderate $\rho$ ($\rho<\rho_\sr(d)$) or dominates at larger $\rho$ ($\rho>\rho_\sr(d)$).
The computation of $\rho_\sr(d)$ is not accessible perturbatively, but  is approximated by JFT by a linear interpolation 
between the exact result  $(\rho, d) = (1/4,1)$ and the point $(\rho, d) = (1,4)$,
deduced from a  mapping to the Burgers equation with non-conserved noise (however, see below).
Above $d_c(\rho)$, the two phases, smooth and rough,  exist and JFT find that the LR noise is always relevant in the smooth phase
while it is always irrelevant in the rough phase.
From their results, they infer that the upper critical dimensions 
of the roughening transition and of the SR rough phase below $d_c(\rho)$ are $d=4$. JFT also conjecture 
that the SR rough phases above and below $d_c(\rho)$ may be of two different natures (called SR-I and SR-II in their paper), 
with possibly  different upper critical dimensions  \cite{Janssen99,Frey99}.

In the present paper, we revisit the  work by JFT using the nonperturbative renormalization group (NPRG) approach, 
successfully developed for the (SR  noise) KPZ equation \cite{Canet10,Canet11a,*Canet12Err,Kloss12}, and here
generalized  to include Gaussian LR correlated noise. 
We derive the corresponding NPRG flow equations at the Next-to-Leading Order (NLO) approximation of Ref.\ \cite{Kloss12}, and 
 solve them to determine  the full phase diagram of the system for various values of $\rho$ and $d$. 
 Our results are in close agreement with the results of JFT in the weak-coupling sector. We recover in particular the 
 smooth LR phases predicted above $d_c(\rho)$ with their exact critical exponents and correction-to-scaling exponents. Furthermore, we find the two stable fixed 
 point  solutions in the strong-coupling regime (in their respective existence domain),  describing  the SR and the LR rough phases,  with the exact LR exponents, 
 and we compute the stability boundary line $\rho_\sr(d)$.
The obtention of  the complete phase diagram of the system in the $(\rho,d)$ plane with all the expected fixed points constitutes our main result.
 In particular, we find that there exists a unique strong-coupling fixed-point
 describing the SR rough phase in all dimensions, which is not consistent with the conjecture by JFT of the existence of two 
 different rough phases SR-I and SR-II above and below $d_c(\rho)$.
 Furthermore, we investigate the phase diagram in the strong-coupling regime around $d = 4$, at least qualitatively since the NLO approximation is no longer accurate in this regime for  $d \gtrsim 3.5$. Combining our findings  and critical exponents
 from numerical simulations \cite{Ala-Nissila93,Castellano98,Castellano99,Marinari00,Pagnani13}, we argue that $d = 4$ may not necessarily be  the upper critical dimension of the 
 SR rough phase. However, as the value of the SR roughness exponent in $d=4$ cannot be reliably determined at this level of approximation, we cannot conclude yet about the actual value of $d_c$ within NPRG, which requires a higher order approximation and is left for future investigation.
 
The remainder of the paper is organized as follows. In Sec.\ \ref{NPRG}, we briefly present the NPRG formalism for 
the KPZ equation, including LR correlated  Gaussian noise, and the approximations used. We then derive the corresponding 
flow equations. These equations are numerically integrated in Sec.\ \ref{RES}, and the full phase diagram of the system is determined and presented, including a discussion about the upper critical dimension.

\section{Nonperturbative renormalization group}
\label{NPRG}

\subsection{KPZ field theory and symmetries}

The field theory associated with
 the KPZ equation (\ref{eqkpz}) with both SR noise and Gaussian LR correlated noise is derived in 
 Ref.\ \cite{Janssen99}, following  the  Janssen-de Dominicis procedure \cite{janssen76, *dominicis76}.
The KPZ dynamic generating functional is given by
\begin{subequations}
\begin{align}
\!\!\!\!{\cal Z}[j,\tj] \! &= \!\!\int \!{\cal D}[h,i \tih]\, 
\exp \! \left(-{\cal S}[h,\tih] +  \int_{\bx} \left\{ j h+\tj\tih\right\} \right) , \label{Z}\\
\!\!\!\!{\cal S}[h,\tih]  \! &= \!\! \int_{\bx}  \!\left\{ \tih(\bx)\left(\p_t h(\bx) -\nu \,\nabla^2 h(\bx) - 
\frac{\lambda}{2}\,\left({\nabla} h(\bx)\right)^2 \right) \right\}\nonumber \\
 & - \!\! \int_{\bq}  \!\left\{  D\, \tih(-\bq)(1+ w q^{-2\rho})\tih(\bq)  \right\}
\label{ftkpz}
\end{align} 
\end{subequations}
where $\tih$ is the Martin-Siggia-Rose response field \cite{Martin73}, $j$ and $\tj$ are sources,
 and the notation $\bx\equiv (t,\vx)$, $\bq\equiv (\omega,\vq)$ was introduced.

The symmetries of the KPZ action with correlated noise are two-fold: (i) the $h$-shift symmetry, (ii) the Galilean symmetry.  
The additional  discrete time reversal symmetry of the one-dimensional SR KPZ equation is no longer realized in presence of 
correlated noise. Moreover, as in the SR KPZ case, the  symmetries (i) and (ii) are gauged in time 
\cite{Canet11a,Lebedev94} and correspond to the following infinitesimal field transformations:
\begin{subequations} 
\begin{eqnarray}
\text{(i)} && \left\{
\begin{array}{l}
h'(t,\vx)=\vx \cdot \p_t \vec v(t) + h(t,\vx+ \lambda \vec v(t))\label{galg}\\
\tilde h'(t,\vx)=\tilde h(t,\vx+ \lambda \vec v(t))
\end{array}
\right.\\
\text{(ii)} && \;\;\;\; h'(t,\vx)=h(t,\vx)+c(t). \label{timeg}
\end{eqnarray}
 \end{subequations}
where $c(t)$ and $\vec v(t)$ are arbitrary infinitesimal time dependent quantities. 
 The  variations of the KPZ action (\ref{ftkpz}) under these time-gauged transformations are linear in the fields, 
 and thus entail simple Ward identities, with a stronger content than  the usual   non-gauged ones  \cite{Canet11a}. 
 The detailed analysis of these  symmetries is at  the heart of the construction of the NPRG approximation 
 scheme, derived in  \cite{Canet11a}.

\subsection{NPRG formalism}

The general NPRG formalism for nonequilibrium systems is presented  in Ref.\  \cite{Canet11b,Berges12,Kopietz10}, 
and its specific  application to the KPZ equation  in Ref.\  \cite{Canet11a}. We only recall here the main elements, following Ref.\ \cite{Canet11a}. 
In the spirit of Wilson's RG ideas, the NPRG formalism  consists in building  a sequence of scale-dependent  effective models   
such that fluctuations are smoothly averaged as the (momentum) scale $\kappa$ is lowered 
from the  microscopic scale $\Lambda$, where no fluctuations are yet included, to the macroscopic scale $\kappa=0$, where they
 are all summed over \cite{Berges02,Delamotte05}.
For classical nonequilibrium problems, one formally proceeds as in equilibrium,
 but with  the presence of the response field, and additional requirements
 stemming from  It$\bar{\rm o}$'s discretization and  causality issues \cite{Canet11b,Benitez12b}. 

To achieve the separation of fluctuation modes within the NPRG procedure, one  adds to the original action ${\cal S}$
a momentum and scale dependent mass-like term:
\begin{equation}
\Delta {\cal S}_\kappa \!=\!\frac{1}{2}\! \int_{\bf q}\!  h_i(-{\bf q})\, 
[R_\kappa({\bf q})]_{ij}\, h_j({\bf q}) ,\;\;\label{deltask}
\end{equation}
where the indices $i,j\in\{1,2\}$ label the field and response field, respectively $h_1=h,h_2=\tih$, and 
summation over repeated indices is implicit. 
The matrix elements $[R_\kappa({\bf q})]_{ij}$ are proportional to a cutoff function
 $r(q^2/\kappa^2)$, with $q=|\vq|$, which ensures the selection of 
fluctuation modes: $r(x)$ is required to almost vanish for $x\gtrsim 1 $ such that 
the fluctuation modes $h_i(q \gtrsim \kappa)$ are unaffected by 
$\Delta {\cal S}_\kappa$,
and to be large  when $x\lesssim 1 $ such that the other modes ($h_i(q\lesssim \kappa)$) are essentially frozen. Furthermore,  
$\Delta {\cal S}_\kappa$ must preserve all the symmetries of the problem and causality properties. As advocated in \cite{Canet11a}, an appropriate choice is 
\begin{equation}
R_\kappa(\omega,\vq) \!\equiv\!R_\kappa(\vq) \!=\! r\left(\frac{q^2}{\kappa^2}\right)
\left(\!\! \begin{array}{cc}
0& {\nu_\kappa} q^2\\
{\nu_\kappa} q^2 & -2 D_\kappa 
\end{array}\!\!\right) \;,
\label{Rk}
\end{equation}
where the running coefficients $\nu_\kappa$ and $D_\kappa$, defined later (Eq.\ (\ref{eq:dknukdef})),  are introduced in the regulator for  convenience \cite{Canet10}. 
Here we choose the cutoff function 
\begin{equation}
r(x)=\alpha/(\exp(x) -1). 
\label{eq:expReg}
\end{equation}
The dependence of our results on the parameter $\alpha$ is discussed in Appendix B.

In the presence of the mass term $\Delta {\cal S}_\kappa$, the generating functional (\ref{Z}) becomes scale dependent 
\begin{equation}
{\cal Z}_\kappa[j,\tj] \!\! = \!\!\!\int {\cal D}[h,i \tih]\, 
\exp\left(-{\cal S}-\Delta{\cal S}_\kappa+  \int_{\bf x} \left\{ j h+\tj\tih \right\}\right) . \label{Zk}
\end{equation} 
  Field expectation values in the presence of the external sources $j$ and $\tilde{j}$ are obtained from the functional  ${\cal W}_\kappa = \log {\cal Z}_\kappa$ as 
\begin{equation}
  \varphi({\bf x}) = \langle h({\bf x}) \rangle = \frac{\delta {\cal W}_{\kappa}}{\delta j({\bf x})}  \, \, , \, \,
  \tilde \varphi({\bf x}) = \langle \tilde h({\bf x}) \rangle = \frac{\delta {\cal W}_{\kappa}}{\delta \tilde j({\bf x})}  .
\end{equation} 
The effective action $\Gamma_\kappa[\varphi,\tilde\varphi]$ is defined as the Legendre transform of  ${\cal W}_\kappa$
(up to a term proportional to $R_\kappa$) \cite{Berges02,Delamotte12,*delamotte07,Canet11b}:
\begin{equation}
\Gamma_\kappa[\varphi,\tilde\varphi] +{\cal W}_\kappa[j,\tj] = 
\int \! j_i \varphi_i -\frac{1}{2} \int \varphi_i \, [R_\kappa  ]_{ij}\, \varphi_{j} .
\label{legendre}
\end{equation}
The exact flow for $\Gamma_{\kappa}[\varphi,\tilde\varphi]$ is given  by Wetterich's 
equation, which  writes in Fourier space  \cite{Wetterich93,Berges02}
\begin{equation}
\partial_\kappa \Gamma_\kappa = \frac{1}{2}\, {\rm Tr}\! \int_{\bf q}\! \partial_\kappa R_\kappa \cdot G_\kappa ,
\label{eq:dkgam}
\end{equation}
where
\begin{equation}
 G_\kappa=\left[\Gamma_\kappa^{(2)}+R_\kappa\right]^{-1}
 \label{eq:propag}
\end{equation}
is the full, that is, field-dependent, renormalized propagator of the theory.
When $\kappa$ is lowered from $\Lambda$ to 0, $\Gamma_\kappa$ interpolates 
between the microscopic model $\Gamma_{\kappa=\Lambda}={\cal S}$  and the full effective action $\Gamma_{\kappa=0}$
 that encompasses all the macroscopic properties of the system \cite{Canet11b}.
Of course Eq.\  (\ref{eq:dkgam})  cannot be solved exactly,  one has to resort to an appropriate   approximation scheme, 
adapted to the specific model under study,  and in particular to its symmetries.

\subsection{Approximations}

\subsubsection{Next-to-Leading Order (NLO) approximation}

In Ref.\ \cite{Canet11a}, inspired by the previous work in equilibrium statistical mechanics of Refs. \cite{blaizot06,benitez09,Benitez12},
an approximation scheme is devised, which consists in building an \anz for $\Gamma_\kappa$  
explicitly preserving the gauged shift (\ref{timeg}) and gauged Galilean (\ref{galg}) symmetries. The building blocks are the Galilean invariants 
$\tilde \varphi$, $\nabla^2 \varphi$, the covariant time derivative $D_t \varphi\equiv \p_t \varphi -(\nabla\varphi)^2/2$, combined with the operators
 $\tilde D_t\equiv  \partial_t- \nabla\varphi\cdot\nabla$ and $\nabla^2$. We  work here in the 
  rescaled theory where $\nu=D=1$ and $\lambda \to \sqrt{g_b} = \lambda D^{1/2}/\nu^{3/2}$. 
Within this scheme, the  `second order' (SO) \anz for $\Gamma_\kappa$  writes 
\begin{align}
\Gamma_\kappa[\varphi,\tilde \varphi]=&   \dis \int_{\bf x} 
\left\{ \tilde \varphi f_\kappa^\lambda(\tm\tilde D_t^2,\tm\nabla^2) D_t\varphi - 
\tilde \varphi f_\kappa^\td(\tm\tilde D_t^2,\tm\nabla^2) \tilde \varphi \right. \nonumber \\
  &\hspace*{-10ex} -\frac 1 2 \left[\nabla^2 \varphi f_\kappa^\nu(\tm\tilde D_t^2,\tm\nabla^2) \tilde \varphi + 
  \left. \tilde \varphi f_\kappa^\nu(\tm\tilde D_t^2,\tm\nabla^2)  \nabla^2 \varphi\right] \right\},
\label{anzso}
\end{align}
where $f_\kappa^\xx$, $X \in\{\nu,D,\lambda\}$ are three running functions.
  It is a truncation  at quadratic order in the response field $\tilde \varphi$, while  the complete momentum and 
  frequency dependence of the 2-point functions is preserved. Note that infinite powers of the field itself 
  are included through the covariant time derivatives $\tilde D_t$. At the bare level $\kappa=\Lambda$, and 
  for purely local noise,  one has $f_\Lambda^\lambda = f_\Lambda^\nu = f_\Lambda^\td =1$, 

The SO flow equations for the functions $f_\kappa^\xx$, derived in \cite{Canet11a}, were integrated  
in the simpler one-dimensional case, where the additional time reversal symmetry imposes that
there remains only one independent running function. The scaling 
functions  associated with the 2-point correlation function were computed, and showed an impressive agreement 
with the exact results \cite{Canet11a,Praehofer04}. However, the integration of the SO flow equations  in generic dimensions  
appears rather involved, and a further simplification was proposed in \cite{Kloss12}. 
 This approximation, referred to as NLO, consists in neglecting the frequency dependence of the three flowing 
 functions $f_\kappa^\xx(\omega,\vp)\to f_\kappa^\xx(\vp)$ within the loop integrals, that is, the right hand side of the flow equations. The NLO flow equations can be 
 found in \cite{Kloss12}, where they were integrated in $d=2$ and $d=3$ and  the scaling functions associated with 
 the 2-point correlation and response functions were computed. The related prediction for a universal amplitude 
 ratio in $d=2$ was very recently confirmed with great accuracy in lattice simulations \cite{Halpin-Healy13}.

In the present paper, we work with LR noise at the NLO approximation. Moreover, we  
focus on zero external frequency,  since we are merely interested in the phase diagram and in the critical 
exponents, and not in the full scaling functions. The zero-frequency sector is decoupled from the non-vanishing 
frequency sector within the NLO approximation, and we denote $f^\xx_\kappa(\omega=0,\vp)\equiv f^\xx_\kappa(\vp)$ for simplicity.
 The inclusion of the noise Eq.\ (\ref{eq:lr-noise-fourier1}) then simply amounts to the substitution 
\begin{equation}
f_\kappa^\td(\vp) \to {\cal D}_\kappa(\vp) = f_\kappa^\td(\vp) + w_\kappa p^{-2\rho}
\label{sub}
\end{equation}
(with bare condition $w_\Lambda = w$) in the NLO \anz.
The first term $f_\kappa^\td(\vp)$ corresponds to the renormalized SR contribution at scale $\kappa$, and the second term 
to the LR one. This separation in terms of a regular  ($f_\kappa^\td(\vp)$) and a nonanalytic part ($w_\kappa p^{-2\rho}$)
holds for any $\kappa$ because, as the flow is regularized in the IR and finite in the UV, it cannot generate nonanalytic contributions.
Correspondingly, the nonanalytic part is not renormalized ($\partial_\kappa w_\kappa = 0$)
and the coupling $w_\kappa$ remains equal to its bare value. Thus, in the presence of LR noise, 
the NLO flow equations for the three functions $f_\kappa^\xx$ are identical to those for the local SR case, up to the 
substitution (\ref{sub}).

The gauged shift symmetry implies the non-renormalization of $f_\kappa^\lambda(0)$ that therefore remains equal to unity for all $\kappa$.
Moreover, the Galilean symmetry implies the non-renormalization of the non-linear coupling $\lambda$.
We hence define two scale dependent parameters $D_\kappa$ and $\nu_\kappa$  
\begin{equation}
 D_\kappa \equiv f_\kappa^\td(0) , \quad
\nu_\kappa \equiv f_\kappa^\nu(0).
\label{eq:dknukdef}
\end{equation}
 These two running coefficients yield  two running anomalous dimensions, defined according to
 \begin{equation} 
 \etad_\kappa = -\kappa \partial_\kappa \ln D_\kappa    \quad \hbox{and}\quad
 \etan_\kappa = -\kappa\partial_\kappa\ln \nu_\kappa ,
  \label{eq:etaflow}  
  \end{equation}
which fixed point values, indexed by *, are related to the physical critical exponents by
\be
z=2-\etan_*  \quad , \quad \chi = (2-d+\etad_*-\etan_*)/2. 
\label{eq:expo}
\ee

In order to study fixed point properties, we introduce  dimensionless quantities. The dimensionless couplings are
\begin{subequations}
\begin{align}
 \hat{w}_\kappa &= {w}_\kappa\,D_\kappa^{-1} \kappa^{-2\rho}, \\
\tg_\kappa &= g_b\,\kappa^{d-2}\, D_\kappa/\nu_\kappa^3, 
\end{align}
\end{subequations}
and their flow equations, due to the non-renormalization of ${w}_\kappa$ and $\lambda$, are hence reduced to their dimensional parts
\begin{subequations}
\begin{align}
 \partial_s \hat{w}_\kappa &= \hat{w}_\kappa (\etad_\kappa - 2\rho), \\
 \partial_s \tg_\kappa &= \tg_\kappa (d-2+3\etan_\kappa-\etad_\kappa).
\end{align}
\label{eq:uflow}
\end{subequations} 
with $\partial_s \equiv \kappa \partial_\kappa$.
 The dimensionless running functions are defined by
 \begin{equation}  
 \tf_\kappa^\xx(\tp)  = f_\kappa^\xx(p)/X_\kappa
  \label{eq:dimlessFunc}  
 \end{equation}  
for $X \in \{D,\nu,\lambda\}$ and  $X_\kappa \in  \{D_\kappa,\nu_\kappa,1\}$,  and their flows write
\begin{align} 
 \partial_s \tf_\kappa^\xx(\tp) 
  &= \etax_\kappa \tf_\kappa^\xx(\tp)+ \tp \;\p_{\tp} \tf_\kappa^\xx(\tp) +\tI_\kappa^\xx(\tp) ,
  \label{eq:dimlessFlowf}
  \end{align}
with $\tp = p / \kappa$, $\eta^\xx_\kappa \in  \{\eta^\td_\kappa,\eta^\nu_\kappa,0\}$,
and the $\tI_\kappa^\xx(\tp)$ are the loop integrals, 
which explicit expressions are given in Ref.\ \cite{Kloss12}, up to the substitution (\ref{sub}).

The five flow equations (\ref{eq:uflow},\ref{eq:dimlessFlowf}) are solved numerically with Euler time 
stepping and $\Delta s = - 4 \times 10^{-4}$ in the RG ``time'' $s$. The three flowing functions $\tf_\kappa^\xx$ 
are set to unity at the initial scale $s=0$. We observe that the flow always converges to a stable 
fixed point, which nature depends on the initial conditions for $\tg_\Lambda=g_b$ and $\hat{w}_\Lambda$. 
  From these flows, one then deduces the phase diagram in the $(\tg,\hat{w})$ 
plane  for each value of the parameters $(\rho, d)$, which is discussed in Sec.\ \ref{RES}.

\subsubsection{Local potential approximation}

As studied in detail in Ref.\ \cite{Kloss12}, the NLO approximation gives a reliable quantitative description of the SR fixed point up to $d \simeq$ 3.5. 
However, the numerical cost to solve the coupled NPRG flow equations is  high, especially as  the flow, in the 
vicinity of unstable fixed points, slows down to impractical timescale. To fully explore the phase diagram, it is 
therefore convenient to sometimes resort to  an additional approximation, usually referred to as the Local 
Potential Approximation prime  (LPA') \cite{Delamotte12}
where  only  field-independent  renormalization coefficients are kept. It thus consists in the following simplification:
\begin{equation}  
 \tf_\kappa^\xx(\tp)  \rightarrow \tf_\kappa^\xx(0) \equiv 1 .
  \label{eq:simple}  
\end{equation}  
The LPA' was shown to already capture the qualitative structure of the phase diagram in the pure 
SR case, although the estimate for the critical exponents rapidly deteriorates as the dimension grows \cite{Canet05b}. 
This approximation will be used to determine the weak-coupling part of the phase diagram.
The complete NLO approximation is however necessary to study the boundary between the SR and LR dominated rough phases 
in $d =$ 2 and 3.  It indeed turns out that the value 
of the roughness exponent $\chi$  is overestimated at the LPA',  such that the stability change of the SR and LR fixed points 
is shifted to unphysical values where $\rho > d /2$ in this approximation, see  Eq.\ (\ref{eq:lr-noise}). In the following, we will indicate whether the NLO or the LPA' is used.

\subsection{Change of variables}

As found by JFT, the LR weak-coupling fixed points (EWLR1 and EWLR2, see below) describing the smooth phase when it exists 
 have an infinite noise amplitude coordinate $\hat{w}_*=\infty$. It is therefore convenient 
to change variables such that the fixed point coordinates remain finite. We choose the same variables as JFT \cite{Janssen99,Frey99}, namely
\begin{equation}
 \hat{x}_\kappa = \frac{\hat{w}_\kappa}{1+\hat{w}_\kappa} ,\quad 
 \hat{y}_\kappa = \frac{1}{4 \rho} (1+\hat{w}_\kappa)^2 \hat{g}_\kappa ,
\label{eq:rescal}
\end{equation} 
to simplify the comparison.

In terms of the new couplings $\hat{x}$ and $\hat{y}$, the flow equations (\ref{eq:uflow}) become
\begin{subequations}
\begin{align}
\partial_s \hat{x}_\kappa &=  \hat{x}_\kappa (1 - \hat{x}_\kappa) (\etad_\kappa - 2 \rho ) , \\
\partial_s \hat{y}_\kappa &=  \hat{y}_\kappa  (2 \hat{x}_\kappa (\etad_\kappa - 2 \rho) + d-2+3\etan_\kappa-\etad_\kappa) 
\end{align}
\label{eq:rescal2}
\end{subequations}
where we have implicitly assumed that the anomalous dimensions $\eta_\kappa^\xx$, which depend  
on $\hat{g}_\kappa$ and $\hat{w}_\kappa$, are now expressed in terms of $\hat{x}_\kappa$ and $\hat{y}_\kappa$.

Let us finally define the variable $\hat{y}'_\kappa=v_d \hat{y}_\kappa/4$ where $v_d^{-1} = 2^{d-1} \pi^{d/2} \Gamma(d/2)$ is related to integration volume, which is used for graphical convenience in all the representations  of flow diagrams, Figs.\ \ref{fig:2d-flow}, \ref{fig:3d-flow}, \ref{fig:3d-flow-pert} and \ref{fig:4d-flow}.

\section{Results}
\label{RES}

\subsection{Fixed points}

We study in the following the existence and stability of the fixed point solutions of the NPRG flow equations 
(\ref{eq:rescal2},\ref{eq:dimlessFlowf}) as functions of $d$ and $\rho$.
Three fixed points correspond to a vanishing $\hat{g}_*$  and are thus referred to as Edwards-Wilkinson fixed points. One (denoted EW)
is at $(\hat{x}_*=0,\hat{y}_*=0)$ while two others (EWLR1 and EWLR2) correspond to infinite $\hat{w}_*$, 
that is, $\hat{x}_*=1$. Another fixed point, denoted T (for 
transition) exists at $\hat{x}_*=0$ and $\hat{y}_*>0$ (for $d>2$) and separates at vanishing LR noise amplitude the smooth and rough phases.
All these four fixed points were found perturbatively and their coordinates, stability and associated exponents were obtained exactly
in the Cole-Hopf representation of the theory \cite{Janssen99}. Besides these fixed points, we find three others. Two, denoted SR and LR, describe 
the rough phase, respectively when the LR noise is irrelevant and relevant. 
These fixed points are genuinely nonperturbative, that is, are not accessible at any order of the perturbative expansion. 
The last fixed point, denoted TLR (for transition in the presence of LR noise), exists in a (narrow) band of the $(\rho,d)$ plane that separates, above the band, a region where there exists a transition between the smooth and rough phases,  and below the band,  
a region where there is no stable smooth phase and where the long distance physics is described by either SR or LR (when, initially, 
the amplitude of the noise is nonvanishing).

\subsubsection{Edwards-Wilkinson fixed point}

The EW fixed point corresponds to $(\hat{x}_*,\hat{y}_*) = (0,0)$ which implies $\eta_*^{\td} =  \eta_*^{\nu} = 0$ and  
$\chi_{\ew} = (2 - d)/2 , \quad  z_{\ew} = 2$. This fixed point is always unstable with respect to LR noise, {\it i.e.} in the $\hat{x}$ direction. It is repulsive (resp.\ attractive)
in the $\hat{g}$ (or $\hat{y}$) direction for $d\le 2$ (resp.\ $d> 2$).

\subsubsection{EWLR1 fixed point}
This fixed point is located at $(\hat{x}_*,\hat{y}_*) = (1,0)$. It is always attractive in the $\hat{x}$-direction while it is attractive
(resp.\ repulsive) for $d>d_{\ewlr}(\rho)=2(1+2\rho)$ (resp.\ for $d<d_{\ewlr}(\rho)$) in the $\hat{y}$-direction, see the Appendix A.
At this fixed point, $\eta_*^\nu=\eta_*^\td = 0$, and the exponents are $\chi_{\ewlr} = (2 - d + 2 \rho)/2$ and   $z_{\ewlr} = 2$. The associated correction-to-scaling exponents are $\omega_1 = 2\rho$ and $\omega_2 =d-2-4\rho$, see Appendix A.

\subsubsection{EWLR2 fixed point}
\label{sec:ewlr2}

The EWLR2 fixed point exists for $d\le d_{\ewlr}(\rho)$ (it coincides with EWLR1 at $d=d_{\ewlr}(\rho)$) and is located at $\hat{x}_* = 1$ and 
$\hat{y}_*\ge 0$, see
Eq.\ (\ref{eq:ewlr2pos}). It is always attractive in the $\hat{y}$-direction, while its stability  in the $\hat{x}$-direction 
changes at $d=d_c(\rho)=2(1+\rho)$, from unstable for $d<d_c(\rho)$ to stable for $d>d_c(\rho)$. 
The critical exponents  $\chi_{\ewlr}$ and $z_{\ewlr}$ are identical in the two LR smooth phases.
However, as already emphasized by JFT, they differ by their
correction-to-scaling exponents, which are for EWLR2  $\omega_1 =d-2- 2\rho$ and $\omega_2 =4\rho -(d-2)$,   see Appendix A.

\subsubsection{Transition fixed point}
\label{sectionTLR}

The transition fixed point T exists for $d\geq 2$ at $\hat{x}_{*} = 0$ and $\hat{y}_{*}\geq 0$ (it coincides with EW in $d=2$). 
It is always unstable in the $\hat{y}$-direction.
 In the Cole-Hopf representation, the change of stability of T in the $\hat{x}$-direction occurs exactly at $d=d_c(\rho)$ (or equivalently at $\rho_c(d)=(d-2)/2$),
 simultaneously with the change of stability of EWLR2, {\it via} the appearance of  
  a line of fixed points joining the two fixed points \cite{Janssen99}. 
 Within our approximations, we find, at fixed $d$, that T is stable in the $\hat{x}$-direction at small $\rho$ and that
its stability changes at $\rho_c^T(d)=(d-2+3\chi_{\gt})/2 \lesssim \rho_c(d)$ (or equivalently at $d_c^T(\rho)$) since the exact value for the critical exponent at the transition $\chi_{\gt}=0$ 
  is only recovered approximately within our approximations. (Given that NLO and LPA' are exact at one loop, we find, as expected $\chi_{\gt}=\mathcal{O}(\epsilon^2)$. However, when  $d$ grows, it becomes slightly negative rather than strictly vanishing).

\subsubsection{TLR fixed point}
\label{sec:tlr}
This fixed point is found for $\rho_c^T(d) < \rho < \rho_{c}(d)$  (equivalently, for $d_c^T(\rho)>d>d_c(\rho)$) 
and has coordinates $\hat{x}_{*} > 0$ and $\hat{y}_{*} > 0$. 
It is unstable in the $\hat{y}$-direction and drives the transition between SR and EWLR2.
As explained above, in the Cole-Hopf representation, the stabilities of the two fixed points T and EWLR2 are switched together through the appearance
at  $\rho=\rho_c(d)$ of a fixed line joining them. This feature is not preserved by our approximation. We find instead that at fixed $d$ 
and upon increasing $\rho$, the TLR fixed point first crosses T at $\rho_c^T(d)$, then travels up  the entire 
 plane $0<\hat{x}<1$, before eventually crossing EWLR2 at $\rho_c(d)$. The line of fixed points is thus replaced by
the TLR fixed point which moves very rapidly  between T and EWLR2 as $\rho$ is increased. This feature is probably an 
artifact of our approximations (if the line of fixed points
 is an exact result, valid beyond perturbation theory). However, the flow is modified only in a narrow band between $\rho_c^T(d)$ and $\rho_c(d)$, 
 and the physically observable phases remain unaffected, controlled by the
fully attractive EWLR2 or SR fixed point (compare Figs.\ \ref{fig:3d-flow} and \ref{fig:3d-flow-pert}).

\subsubsection{Short-range fixed point}

The SR fixed point is located $\hat{x}_{*}  = 0$ and $\hat{y}_{*} > 0$. It exists in all dimensions and describes  
the rough (strong-coupling) phase of the KPZ equation without LR noise. Within all our approximations  (except the LPA'), the associated exponents are in  good agreement with the numerical ones in $d=2$ and $d=3$ 
\cite{Canet10,Canet11a,*Canet12Err,Kloss12}. The quality of our
 approximations deteriorates with increasing dimension and none of them yields reliable quantitative results above typically $d=3.5$.
In all dimensions  we find that SR is stable in the $\hat{y}$-direction. Its stability in the noise direction 
 can be inferred from Eq.\ (\ref{eq:expo}) and Eq.\ (\ref{eq:rescal2}), that is
\begin{equation}
 \partial_s \hat{x}_\kappa =  \hat{x}_\kappa (1 - \hat{x}_\kappa) (d - 2 + 3 \chi_{\sr} - 2\rho).
  \label{eq:betauSR}
\end{equation} 
The sign change of the $\beta$-function in the  $\hat{x}$-direction hence occurs  at
\begin{equation}
 d_{\sr} =  2 + 2 \rho - 3 \chi_{\sr}.
 \label{eq:srstability}
\end{equation} 
At fixed $d$, the SR fixed point is attractive in the noise direction for $\rho<\rho_{\sr}(d)=(d-2+3\chi_{\sr})/2$ and becomes repulsive
beyond this value.

\subsubsection{Long-range fixed point}
\label{sec:long-range-fp}

The LR fixed point has coordinates $\hat{x}_{*} > 0$ and $\hat{y}_{*} > 0$ and describes a strong-coupling rough, LR dominated phase.
At fixed $d$, it exists for $\rho>\rho_{\sr}(d)$ (it coincides with SR at $\rho=\rho_{\sr}(d)$) and is attractive in all directions.
Under the assumption that it indeed exists, the  associated exponents have been determined exactly in the Cole-Hopf representation \cite{Janssen99}:
\begin{equation}
 \chi_{\lr} = (2 - d + 2 \rho)/3 ,  \quad z_{\lr} = (4 + d - 2 \rho)/3 
\label{eq:chiLR}
\end{equation} 
and are also obtained exactly from the NPRG Eqs.\ (\ref{eq:expo},\ref{eq:uflow}) at any non-trivial fixed point with nonvanishing LR noise.
Let us notice that the LPA' is not sufficient to find with a reasonable accuracy the value of $\rho_{\sr}(d)$ where
SR and LR exchange their stability and we resorted to the complete NLO approximation to get it.
We compare in Table~\ref{table} the location of the boundary line $\rho_\sr(d)$ between the SR and LR phases obtained with different approaches.
The JFT result corresponds to a linear interpolation $\rho_\sr(d)=d/4$ proposed by these authors \cite{Janssen99}.
 The NLO and the numerical results correspond
to the value of $\rho$ verifying $\chi_{\lr}=\chi_{\sr}$, that is $\rho_\sr(d)=(3\chi_{\sr}+d-2)/2$, where the values for $\chi_{\sr}$ are obtained respectively from the NLO approximation or numerical simulations \cite{Tang92b,Ala-Nissila93,Castellano99,Marinari00,Reis04,Ghaisas06,Kelling11,Pagnani13}.
\begin{table}[h]
\caption{\label{table} Location  of the boundary line $\rho_\sr(d)$ between the SR and LR phases as a function of the dimension $d$ from JFT \cite{Janssen99},  NLO (this work, $\alpha=4$, see Appendix B) and numerical simulations (mean values from \cite{Tang92b,Ala-Nissila93,Castellano99,Marinari00,Reis04,Ghaisas06,Kelling11,Pagnani13}).}
\begin{ruledtabular}
\begin{tabular}{lccccc}
  $d$           & 1   &   2   &   3  & 4    \\ \hline
 $\rho_\sr^{\rm JFT}$ & 1/4 & 1/2   &  3/4 & 1  \\
 $\rho_\sr^{\rm NLO}$ & 1/4 & 0.57  & 0.79 & -- \\
 $\rho_\sr^{\rm num.}$ & 0.25& 0.57  & 0.95 & 1.37
\end{tabular}
\end{ruledtabular}
\end{table}

\begin{figure}[tb]
  \centering
\includegraphics[width=0.5\textwidth]{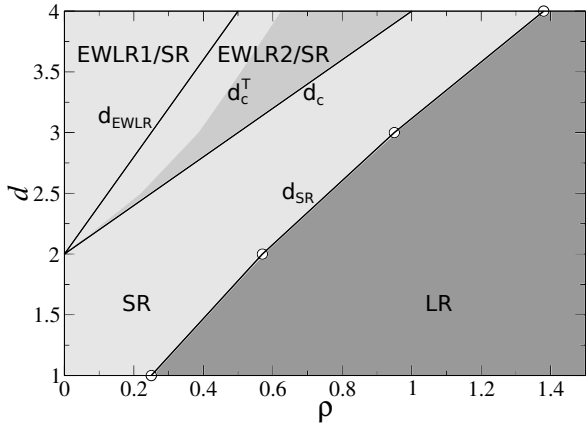}
  \caption{%
Phase diagram of the KPZ equation with spatially LR correlated noise in the ($\rho,d$) plane. Bounds between the  regions, 
which are indicated by  black lines, are given by $d_{\ewlr}(\rho)=2(1+2\rho)$, $d_c(\rho)=2(1+\rho)$ and $d_{\sr}(\rho)$ 
 defined by (\ref{eq:srstability}) where  averaged values for $\chi_{\sr}$ are taken 
from numerical simulations (mean values from \cite{Tang92b,Ala-Nissila93,Castellano99,Marinari00,Reis04,Ghaisas06,Kelling11,Pagnani13}).  LR  is the unique fully
attractive fixed point in the dark gray region. In the other regions there is either a phase transition between a flat and a rough phase
where the LR noise is irrelevant or, below $d_c(\rho)$, only a rough phase described by SR. The TLR fixed point exists in the 
gray region between the lines $d_c^T$ and $d_c$, see \ref{sectionTLR}.
}
\label{fig:phase-diagram}
\end{figure}

\begin{figure}[tb]
 \begin{minipage}{4cm}
 \hspace{-6mm}
\includegraphics[width=45mm]{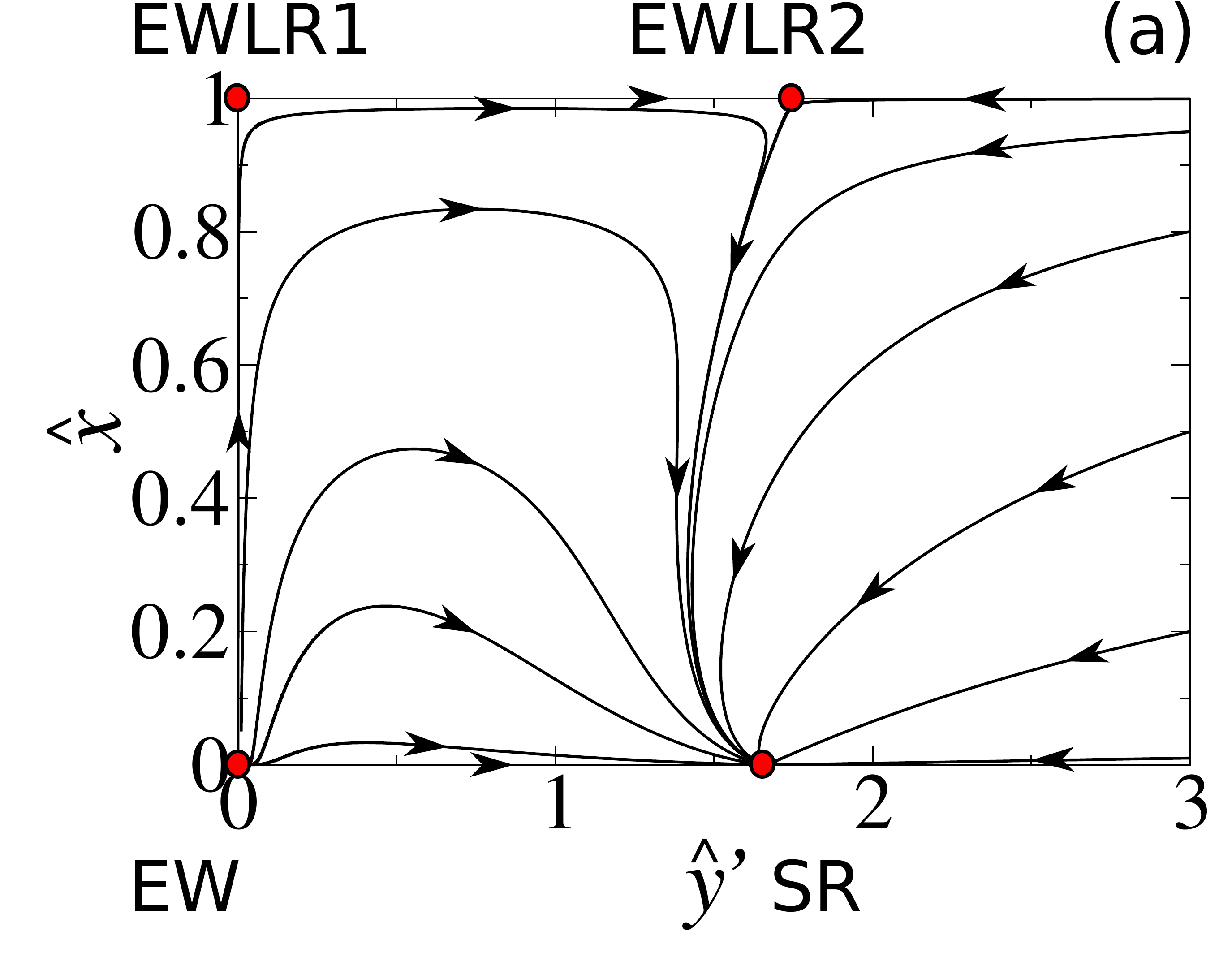}
\end{minipage}
 \begin{minipage}{4cm}
\includegraphics[width=45mm]{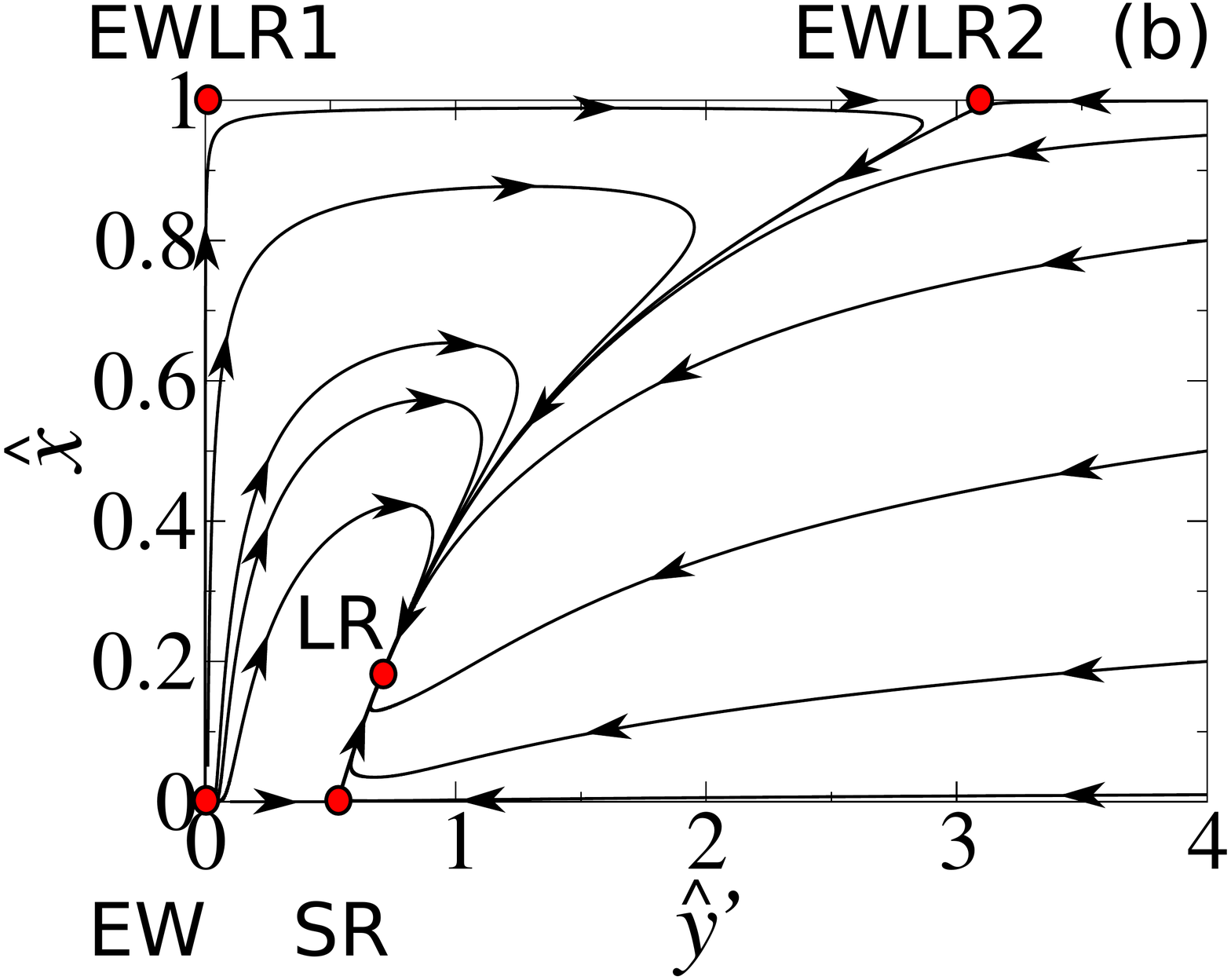}
\end{minipage}
\caption{(Color online) RG trajectories in the $(\hat{y},\hat{x})$ plane in $d=2$ for (a) $\rho=0.3$ and (b) $\rho=0.7$  
obtained with the LPA'. Increasing $\rho$, the LR fixed point moves from the unphysical quadrant $\hat{x}<0$
(not shown in panel (a)) to the physical one (panel (b)) through crossing the SR fixed point
that thus simultaneously changes stability in the $\hat{y}$ direction.}
  \label{fig:2d-flow}
\end{figure}

\subsection{Discussion of the phase diagram}

After having characterized all the fixed point solutions of the NPRG flow equations, we now provide a complete picture of the phase diagram (see Fig.~\ref{fig:phase-diagram}).
There are two distinct situations depending on the dimension. 
 First we confirm the general picture found by JFT. That is, for $d<d_c(\rho)$, the interface is always rough, with a 
 phase boundary $\rho_\sr(d)$ separating the usual strong coupling SR phase for $\rho<\rho_\sr(d)$  and a LR dominated phase with 
 $\rho$-dependent critical exponents for    $\rho>\rho_\sr(d)$. Above $d_c(\rho)$, the T fixed point drives a 
 transition between a smooth LR phase and a rough SR phase.
In the following, we discuss the details of the phase diagram, reasoning rather at fixed $d$ and for varying $\rho$, 
which is closer in spirit to what can be observed in simulations. 

For $d\le2$, the system is always
in a rough phase  and for $\rho\le \rho_{\sr}(d)$ the flow is driven to the SR fixed point whatever the initial condition is, provided 
the nonlinearity is nonvanishing ($\lambda>0$), see Fig.\ \ref{fig:2d-flow} (a). In this case, the presence of 
the LR noise does not change the long distance physics of the KPZ equation. At $\rho=\rho_{\sr}(d)$, the LR fixed point crosses
SR  and enters  the physical quadrant $\hat{x}>0$ for $\rho>\rho_{\sr}(d)$. It is then fully attractive and drives the long distance
physics of any model showing nonvanishing LR noise, see Fig.\ \ref{fig:2d-flow} (b).

\begin{figure}[tb]
 \begin{minipage}{4cm}
 \hspace{-6mm}
\includegraphics[width=45mm]{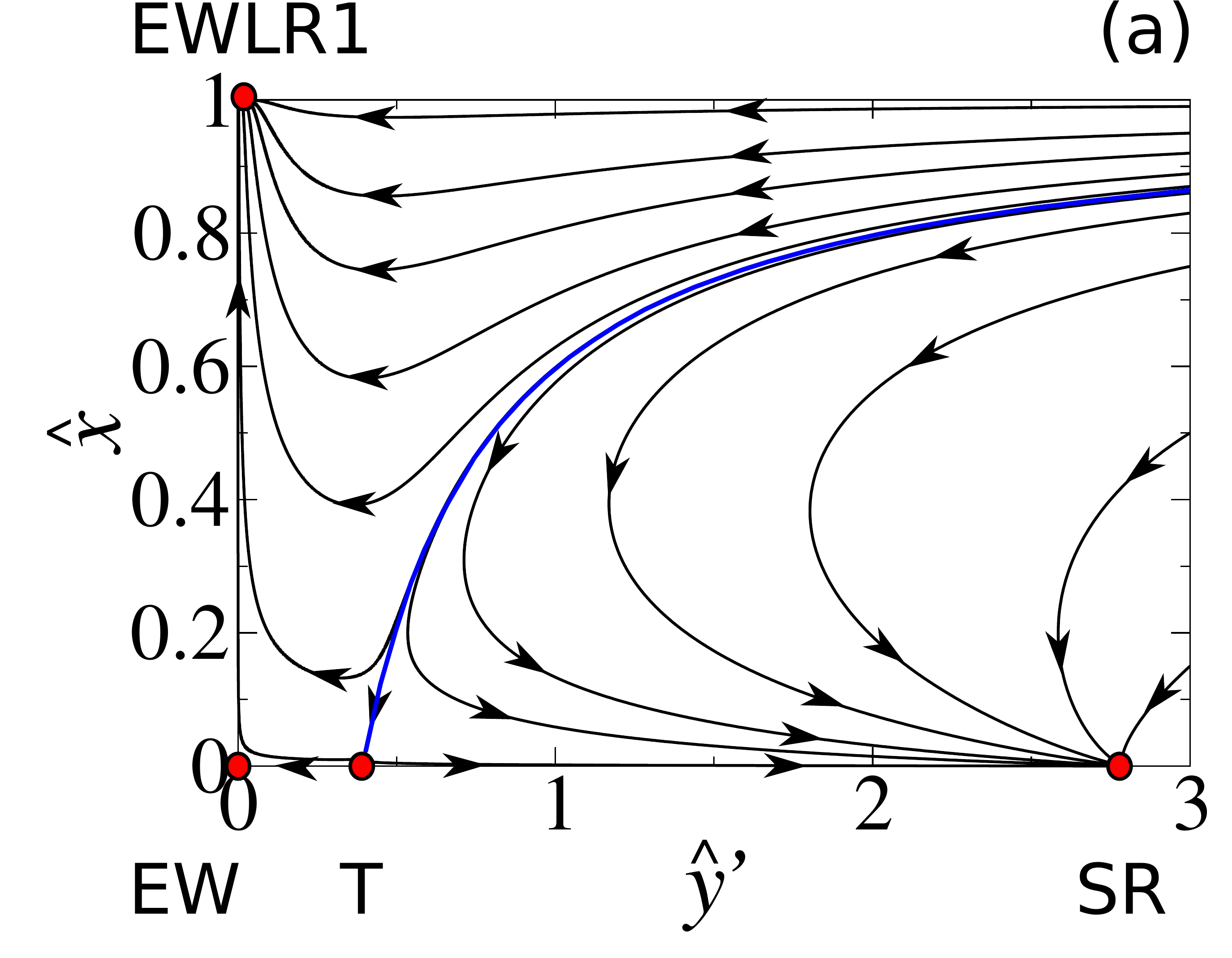}
\end{minipage}
 \begin{minipage}{4cm}
\includegraphics[width=45mm]{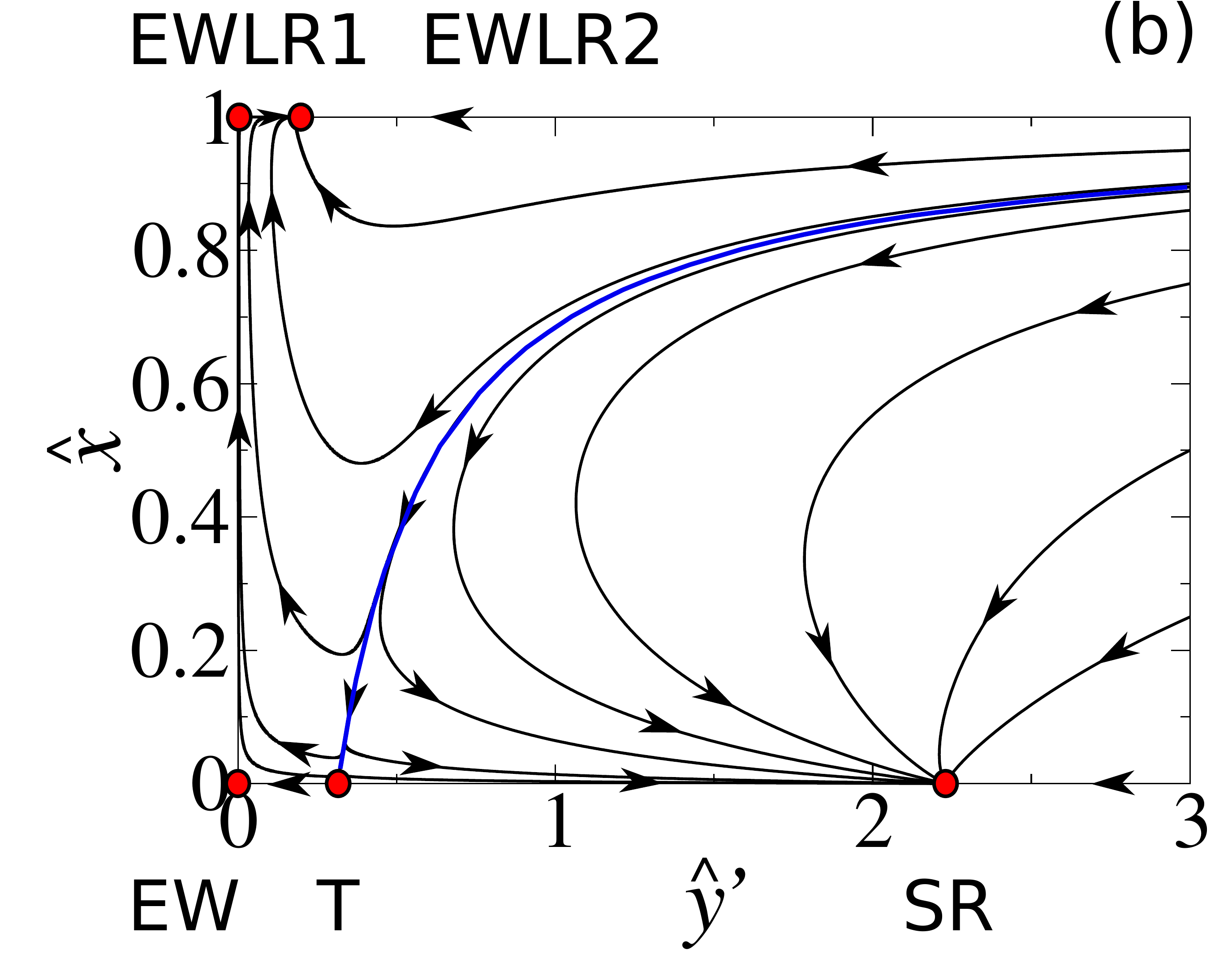}
\end{minipage}
  \vspace{-4mm}
 \begin{minipage}{4cm}
 \hspace{-6mm}
\includegraphics[width=45mm]{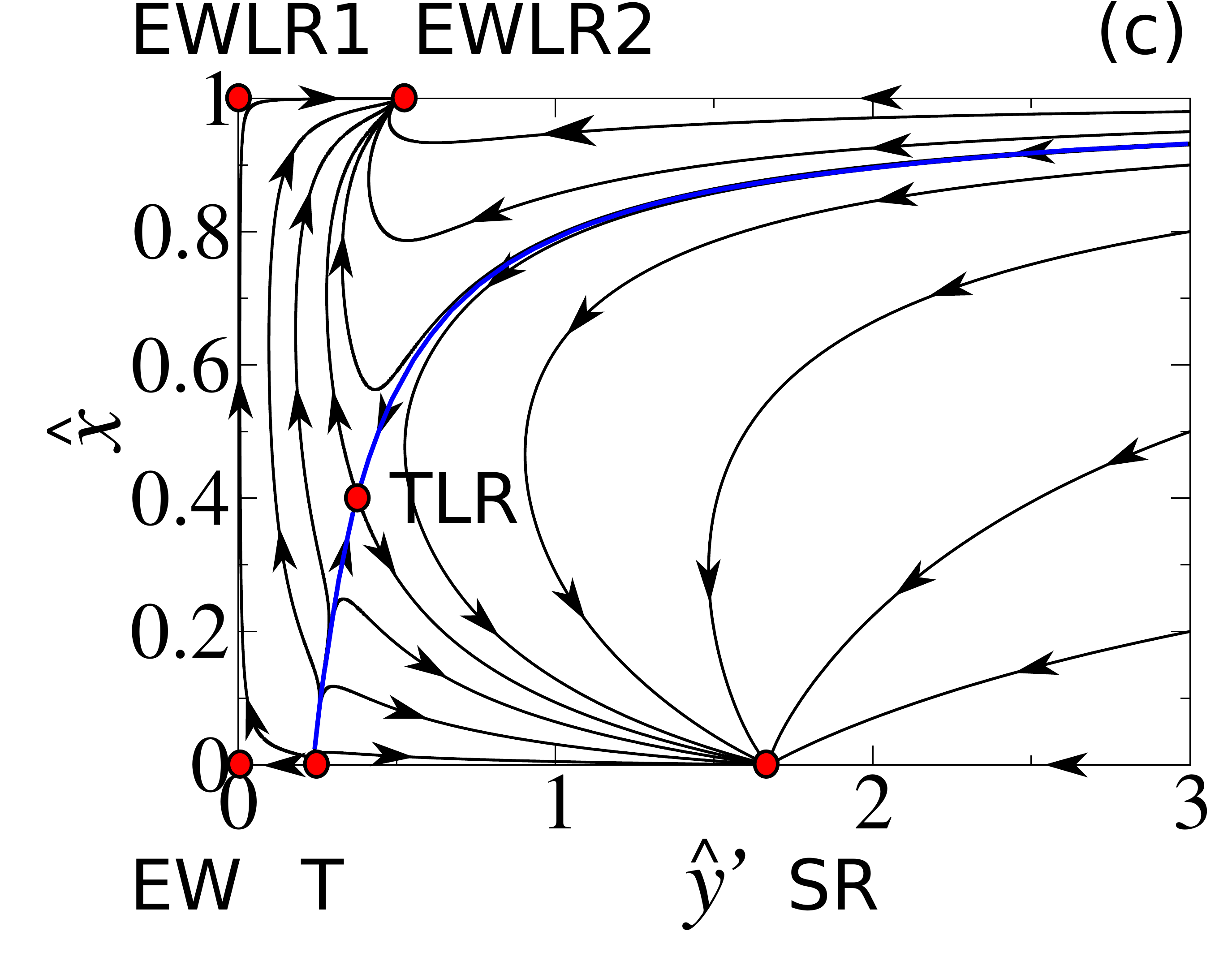}
  \vspace{1mm}
\end{minipage}
 \begin{minipage}{4cm}
\includegraphics[width=45mm]{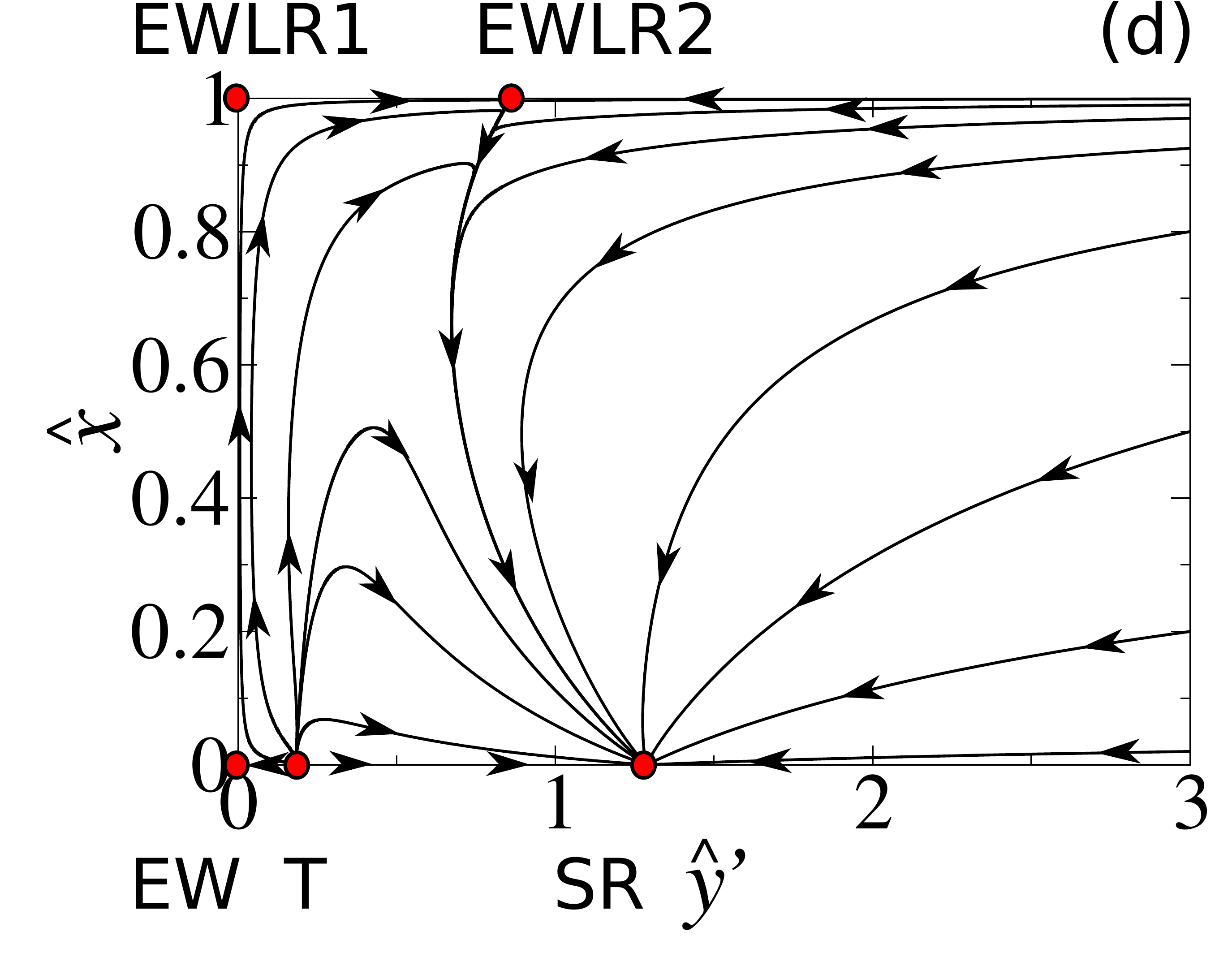}
  \vspace{1mm}
\end{minipage}
\includegraphics[width=45mm]{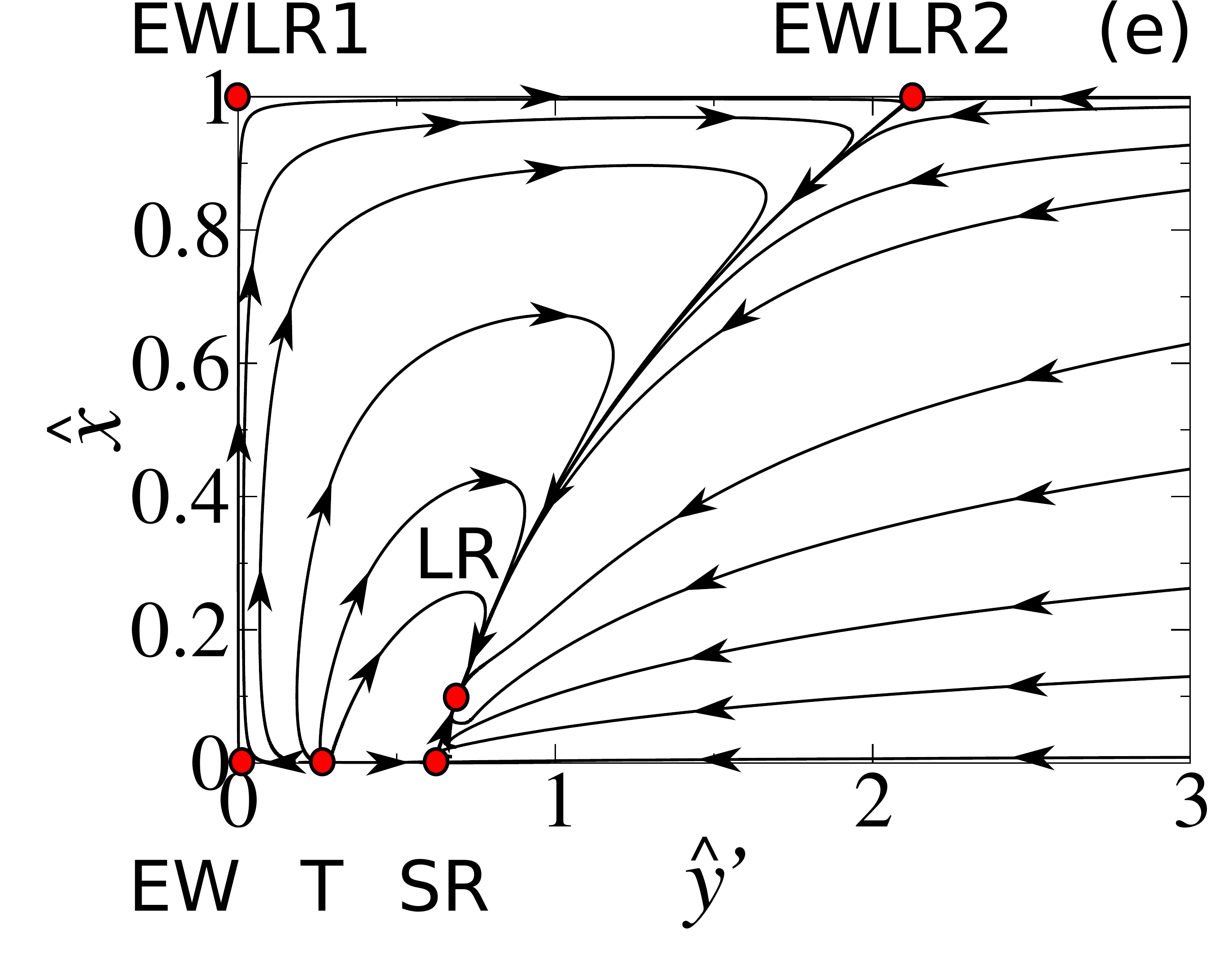}
 \caption{(Color online) RG trajectories in the $(\hat{y},\hat{x})$ plane in $d=3$  for increasing values of $\rho$: 
 (a) $\rho=0.24$, (b) $\rho=0.3$, (c) $\rho=0.4$,  (d) $\rho=0.52$,  (e) $\rho=0.9$, obtained with the LPA' for (a) 
 to (d) and NLO for (e). 
Panels (a) and (b): The EWLR2 fixed point enters into the physical quadrant $\hat{y}'>0$
and EWLR1 changes its stability. Panel (c): The TLR fixed point enters into the physical quadrant
$\hat{x}>0$ and T changes  stability. Panel (d): The
TLR fixed point merges with EWLR2 that changes stability. Panel (e):
The LR fixed point enters into the physical quadrant $\hat{x}>0$
and the SR fixed point changes stability.}
  \label{fig:3d-flow}
\end{figure}

\begin{figure}[tb]
\begin{minipage}{4cm}
 \hspace{-6mm}
 \includegraphics[width=45mm]{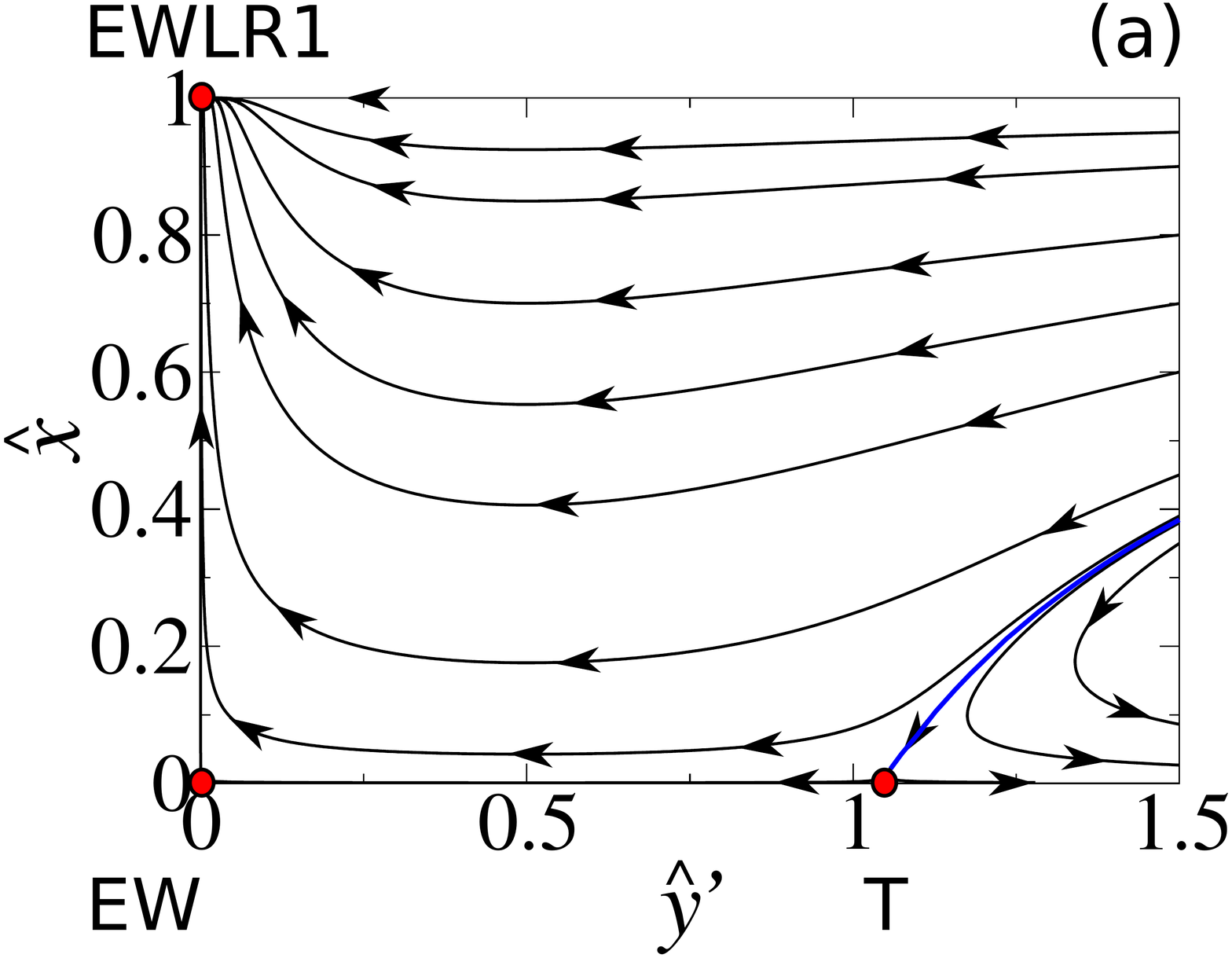}
\end{minipage}
 \begin{minipage}{4cm}
\includegraphics[width=45mm]{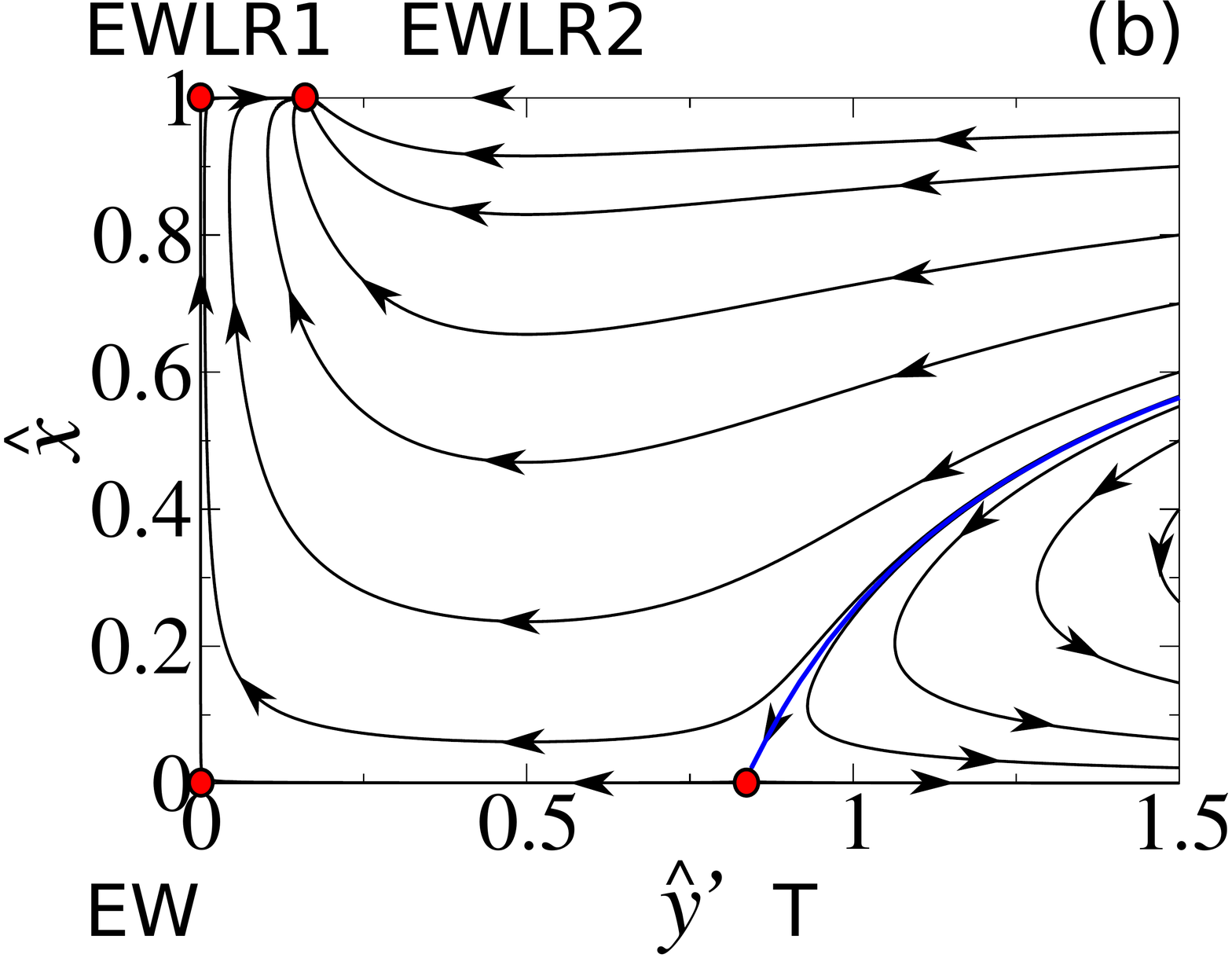}
\end{minipage}
 \begin{minipage}{4cm}
 \hspace{-6mm}
\includegraphics[width=45mm]{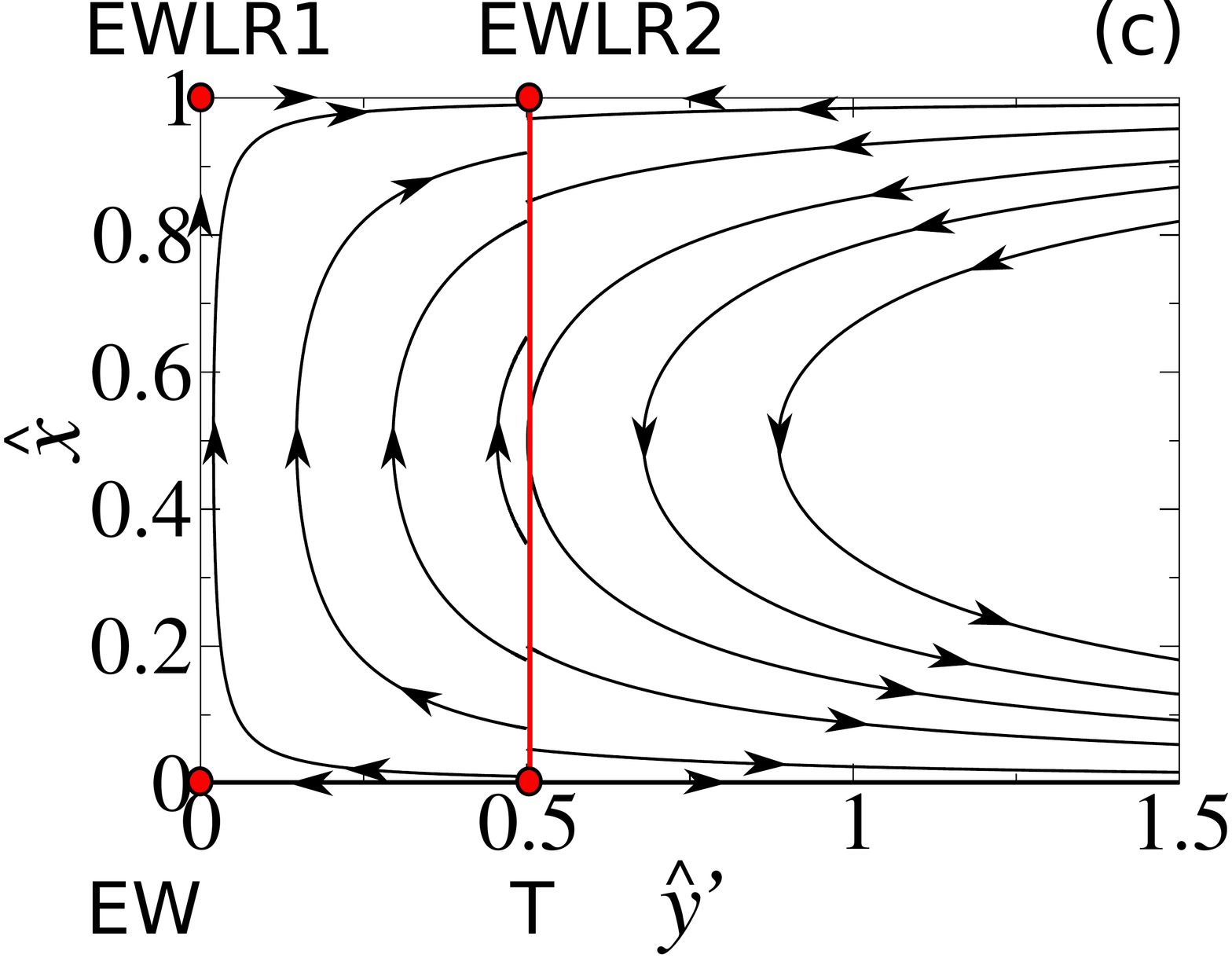}
\end{minipage}
 \vspace{-4mm}
 \begin{minipage}{4cm}
\includegraphics[width=45mm]{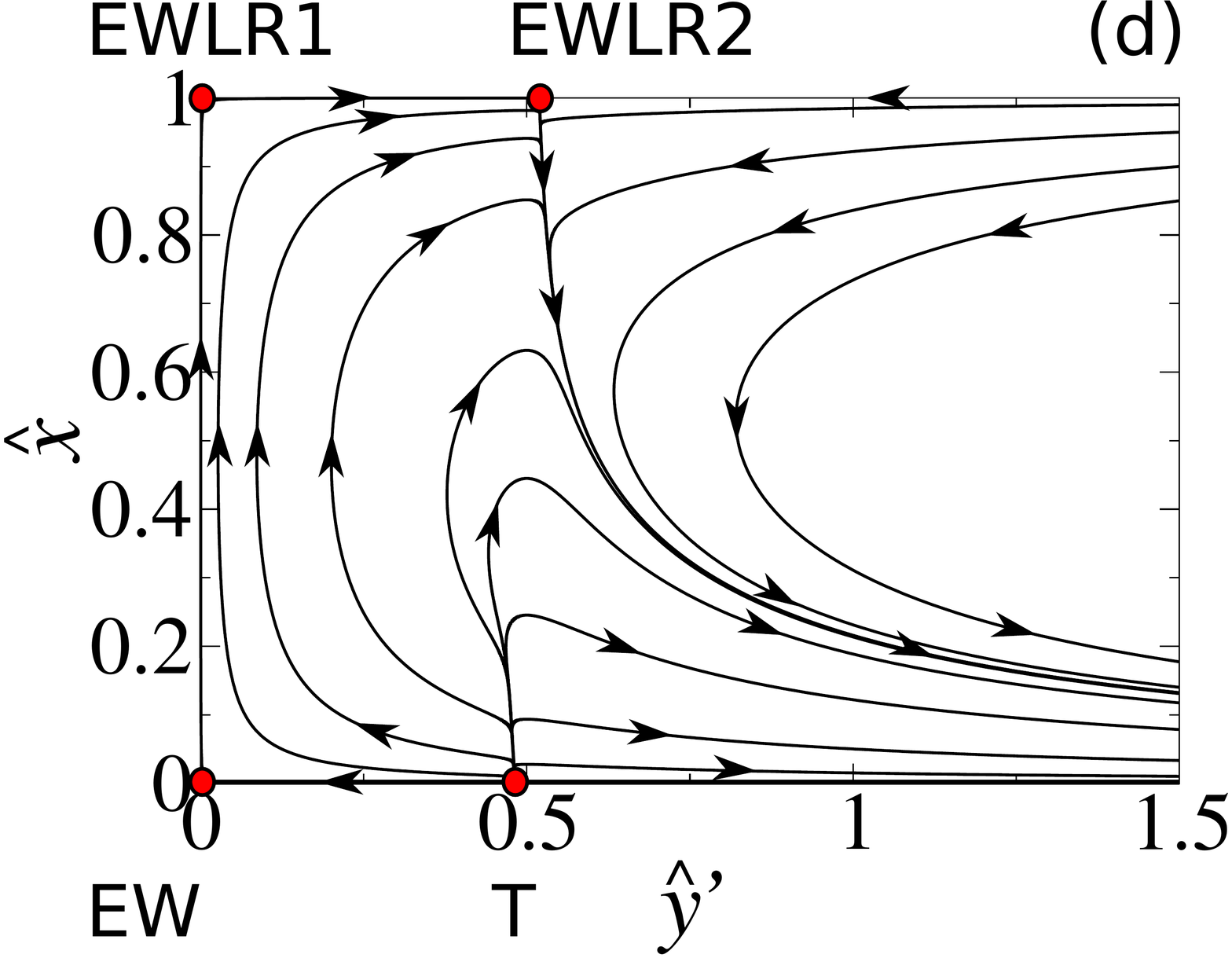}
\end{minipage}
  \caption{(Color online) RG trajectories in the $(\hat{y},\hat{x})$ plane in $d=3$ and for increasing values of $\rho$: 
  (a) $\rho=0.24$, (b) $\rho=0.3$, (c) $\rho=0.5$, (d) $\rho=0.52$, obtained with the perturbative flow equations (\ref{eq:minimalRG}). 
  The rapid move of TLR in Fig.\ \ref{fig:3d-flow} is replaced by a fixed line joining T and EWLR2 displayed in (c), the rest of the (weak-coupling part of the) 
  flow diagrams being very similar to the non-perturbative ones. (Note the difference of $\hat{y}'$ scale).}
  \label{fig:3d-flow-pert}
\end{figure}

For $d>2$, the situation is more complex. At vanishing LR noise amplitude ($w_\Lambda=0$), the fixed point T separates a smooth 
(at small $g_\Lambda$) and a rough (at large $g_\Lambda$) phase. The smooth phase is described by the usual EW fixed point
and the rough phase by the SR fixed point.
For $\rho<\rho_c(d)=(d-2)/2$ and nonvanishing noise amplitudes, there exists a critical line (highlighted in blue in Figs. \ref{fig:3d-flow} and \ref{fig:3d-flow-pert}) ending at T also separating a smooth 
 and a rough phase. This line is  nontrivial as can be seen on the panels (a) to (c) of Fig.\ \ref{fig:3d-flow}. Depending on $\rho$,
the flow in the smooth phase is either driven, for $\rho<(d-2)/4$ to EWLR1, Fig.\ \ref{fig:3d-flow} (a),  or for $(d-2)/2>\rho>(d-2)/4$ to EWLR2, Fig.\ \ref{fig:3d-flow} (b). 
In the rough phase, the flow is driven to SR for $\rho<\rho_\sr(d)$, Figs.\ \ref{fig:3d-flow} (a) to (d), which    becomes fully attractive 
 in the entire $(\hat{y}',\hat{x})$ plane for $\rho_c(d)<\rho<\rho_\sr(d)$, Figs.\ \ref{fig:3d-flow} (d). 
In this case, the LR noise is irrelevant and the LR fixed point lies in the unphysical quadrant $\hat{x}<0$. For $\rho>\rho_\sr(d)$, the LR fixed point crosses 
SR and appears in the physical quadrant $\hat{x}>0$ becoming the dominant, fully attractive fixed point Fig.\ \ref{fig:3d-flow} (e).
 For $\rho>\rho_c(d)$ and $d>2$, the flow is thus
very similar to what  is found in $d\le2$: either $\rho<\rho_{\sr}(d)$ and SR is fully attractive  or $\rho>\rho_{\sr}(d)$ 
and LR is fully attractive and governs the rough phase. 
Let us emphasize that we can follow continuously all these fixed points in the
$(\rho,d)$ plane and that there are no two distinct SR phases contrary to what was conjectured in \cite{Janssen99}.

As already mentioned, if no approximation were performed and in the perturbative Cole-Hopf
approach, a fixed line joining T to EWLR2 would appear exactly at $\rho=\rho_c(d)$ and for $\rho>\rho_c(d)$ both fixed points
would become unstable, see Fig.\ \ref{fig:3d-flow-pert}. Instead, within our approximations and upon increasing $\rho$ at fixed $d$,
the unstable fixed point TLR crosses T for $\rho =\rho_c^T(d)$, moves
very rapidly towards EWLR2 and finally crosses this fixed point at $\rho=\rho_c(d)$,  Fig.\ \ref{fig:3d-flow} (c) and (d),  changing the stability
of these  fixed points upon crossing them. However, this little discrepancy  does not modify qualitatively the rest of the phase diagram and the physically
observable phases are unaffected (compare Figs.\ \ref{fig:3d-flow} and \ref{fig:3d-flow-pert}).

\subsection{Discussion about the upper critical dimension}

\begin{figure}[tb]
 \begin{minipage}{4cm}
 \hspace{-9mm}
\includegraphics[width=48mm]{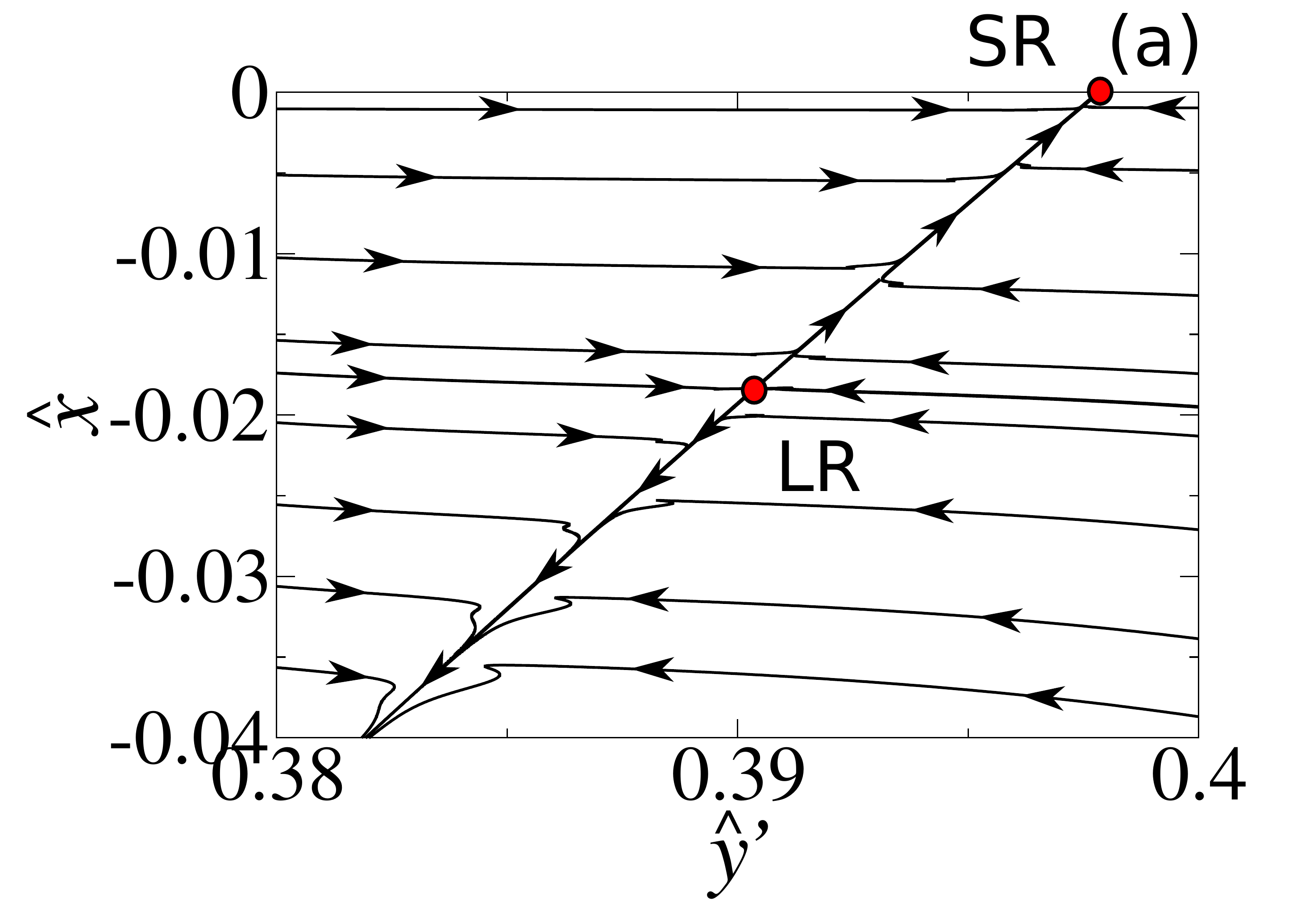}
\end{minipage}
 \begin{minipage}{4cm}
\includegraphics[width=45mm]{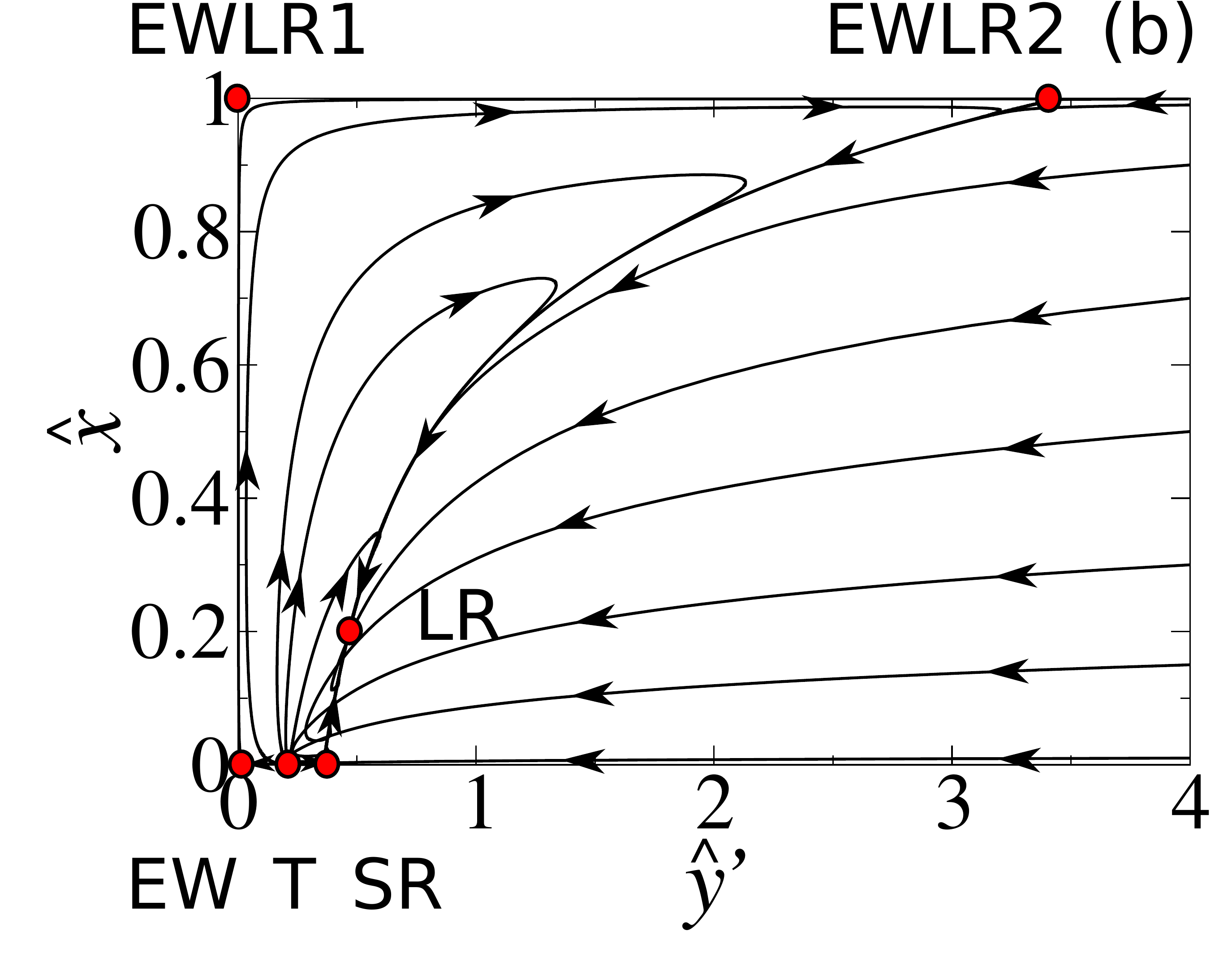}
\end{minipage}
\caption{(Color online) RG trajectories in the $(\hat{y},\hat{x})$ plane in $d=4$ for (a) $\rho=1.125$ and (b) $\rho=1.25$  
obtained with NLO. The LR fixed point lies away from the Gaussian fixed point and is fully attractive for $\rho\gtrsim 1.14$.}
  \label{fig:4d-flow}
\end{figure}

We have followed the LR
fixed point up to dimension 4 for $\rho \simeq 1$ \footnote{In $d=4$, we choose the cutoff parameter $\alpha=10$, see Appendix B.}. 
We observed that it does not lie close to the Gaussian fixed point near $\rho=1$ and $d=4-\epsilon$. In fact, within the NLO approximation,
we find in $d=4$ that a transition between the SR and LR dominated phases occurs at $\rho_\sr(4)\simeq 1.14$, such that the LR fixed point  becomes the stable fixed point for $\rho> \rho_\sr(4)$ with a finite value of $\hat{y}_*$, see Fig.\ \ref{fig:4d-flow}. 
We recall that our results in the strong coupling phase show a large
dependence on the choice of regulators for dimensions larger than typically 3.5, which strongly suggests that our approximations are not accurate in this case, see Appendix B. However, there is no doubt that the LR fixed point
cannot become Gaussian in $d=4$ and $\rho=1$.  As a matter of fact, if it were Gaussian, it would exist as a solution of the perturbative expansion of our NPRG equations since our approximations, either the LPA' or NLO, are exact at one-loop order by construction. There is no such a solution. Put it differently, while the strong coupling regime of the problem becomes out of reach of
 our approximations in large dimensions, the weak coupling regime remains under control.
In the NPRG calculations, the case $d=4$ and $\rho=1$ does {\it not} map onto the Burgers equation with non-conserved noise, ({\it i.e}.\ model B of Forster {\it et al}.\ \cite{Forster77} applied to the Burgers equation). In the latter,   only a LR noise is present and it does not include a SR part, such that nothing can be inferred from  this model about   the  stability of the LR fixed point against a SR component. 
 The reason for the discrepancy with JFT who advocated this mapping is that, in the RG approach, the SR noise is generated
by the flow even if it is not present initially and it cannot be neglected. 
The full complexity of the KPZ equation with both types of noise, and their competition, cannot be avoided
 to determine their respective relevance.
As a result, the usual power counting argument performed in model B (for LR $\rho=1$ without a SR component) leading to an upper critical dimension of 4 for LR cannot be applied here.

Within the NPRG framework, as already mentioned, the NLO approximation is 
 not accurate above $d \simeq$ 3.5.
However, the qualitative structure of the obtained phase diagram in $d=4$ in the strong-coupling sector (Fig. \ref{fig:4d-flow}), together with some inputs from numerical 
 simulations for the critical exponent, open up another possibility, which we now stress.
 The stability exchange between SR and LR proceeds when LR comes across SR from below,  
 which occurs for $\chi_{\sr} = \chi_{\lr}$. Thus, if $\chi_{\sr}>0$
in $d=4$, as  suggested by numerical simulations, SR appears as the stable fixed point for $\rho=1$, 
dominating over  LR, which has $\chi_{\lr}=0$ and  still lies in the unphysical quadrant $\hat{x}<0$ of 
the coupling constant space (see Fig.~\ref{fig:phase-diagram}). From simulation results for $\chi_{\sr}$, the SR stability change in $d = 4$ occurs around $\rho_\sr \simeq 1.38$, and thus SR is still fully attractive  at $\rho = 1$ \cite{Ala-Nissila93,Castellano98,Castellano99,Marinari00,Pagnani13}.

Let us summarize the previous discussions of the phase diagram and of $d=4$. We emphasized that the existence of two different types 
of SR phases above and below $d_c(\rho)$ with different upper critical dimensions is not consistent with our finding of a unique SR fixed point. 
However, we proposed an alternative  scenario which could reconcile numerical simulations and RG analysis, namely that the LR fixed point is 
unstable against SR at $d=4$ and $\rho=1$.  Let us emphasize once more that we cannot settle yet whether this scenario is realized within NPRG, 
{\it i.e}.\ whether $\chi_{\sr}$ is vanishing or not in $d=4$.
 Doing so requires a higher-order approximation, and this issue  will be addressed in future work.

\section{Conclusion}
\label{CCL}

In the present work, we  investigated, using NPRG, the phase diagram of the KPZ equation
 with Gaussian LR correlated noise with power law decaying correlator
 ${\cal D}(p) = D(1+w p^{-2\rho})$ in Fourier space. We generalized the NPRG flow equations in the NLO approximation to include LR noise. 
 We then integrated them numerically to
determine the complete phase diagram of this model as a function of $d$ and $\rho$, and  confronted it with the results obtained  by JFT, 
which are valid to all order in perturbation theory. 

In the weak-coupling sector, the two approaches are in close agreement.
  We recover in particular that above $d_c(\rho)=2(1+\rho)$, the smooth phase is LR dominated and is controlled by one of the two 
  weak-coupling LR fixed points EWLR1 or EWLR2, with their exact critical exponents and correction-to-scaling exponents. One difference 
  appears between the two approaches: the line of fixed point joining T and EWLR2 at exactly $\rho_c(d) = (d-2)/2$ is replaced in the NPRG 
  approach at NLO by an unstable fixed point TLR rapidly moving from T to EWLR2 as $\rho$ is increased in the vicinity of $\rho_c$. 
  This difference originates in the fact that the transition fixed point T is only approximately described within the NLO approximation. 
  It has however a negligible impact on the structure of the phase diagram. 

In the strong-coupling sector, we find the two fixed points that govern the SR and LR rough phases, which constitutes our main result.
They exchange their stability when LR comes across SR from an unphysical quadrant of the coupling space which occurs for $z_{\lr} = z_{\sr}$. 
 We hence computed the phase boundary $\rho_\sr(d)$ which is not accessible within perturbation theory.
All of the fixed points can be followed continuously when $\rho$ and $d$ are varied, and we show in particular that there exists a unique SR fixed point.
This is not consistent with the scenario proposed by JFT of two  SR phases, below and above $d_c(\rho)$ respectively, of different natures.
 We finally 
 suggest that the RG finding does not in fact rule out the possibility that the SR phase (SR-I in JFT's work)  has an upper critical 
 dimension different from 4, and possibly infinite, which would be compatible with the numerical results. 
However, the NLO approximation we used does not allow us to determine whether this possibility is realized within NPRG,
as the NLO approximation does not allow us to  accurately investigate the  $d=4$ case. This is left for future work.

\section{Acknowledgments}

 The authors acknowledges financial support from the ECOS-Sud France-Uruguay program U11E01.
 TK thanks the LPTMC at the Universit\'e Pierre et Marie Curie  for hospitality, where parts of this work were achieved.
TK also gratefully acknowledges financial support from the Alexander von Humboldt foundation, the IIP in Natal and for
hospitality and computing resources in the Group of P. Kopietz
at the Goethe University Frankfurt. Finally, LC and BD thank the Universidad de la Rep\'ublica (Uruguay) for hospitality
and  the PEDECIBA for financial support during the completion of this work.  

\begin{appendix}

\renewcommand{\theequation}{A\arabic{equation}}
\section*{Appendix A: perturbative analysis of NPRG equations}
\setcounter{equation}{0}

In this Appendix, we analyze the NPRG flow equations in some perturbative regimes. In the vicinity of $d=2$ and $\rho=0$, the NPRG flow equations coincide with
Eqs.\ (4.33,4.34) of Ref.\ \cite{Janssen99}, which read in our normalizations, that is with $u_{\rm JFT}= v_d \hat{g}_\kappa /2$:
\begin{subequations}
\begin{align}
\partial_\kappa\hat{g}_\kappa &=  \hat{g}_\kappa \left[\epsilon - v_d \frac{\hat{g}_\kappa}{4} (1+\hat{w}_\kappa)^2\right] , \\
\partial_\kappa \hat{w}_\kappa &=  \hat{w}_\kappa  \left[- 2 \rho + v_d \frac{\hat{g}_\kappa}{4} (1+\hat{w}_\kappa)^2\right] , 
\end{align}
\label{eq:minimalRG}
\end{subequations}
with $\epsilon = d - 2$.

The NPRG $\beta$-functions for these couplings are given by  Eqs.\ (\ref{eq:uflow}).  The  anomalous dimensions are defined at 
zero external momentum, through the normalization conditions $\tf_\kappa^\nu(0)=\tf_\kappa^\td(0)=1$ ensuing from definitions 
(\ref{eq:dknukdef},\ref{eq:dimlessFunc}). Eq.\ (\ref{eq:dimlessFlowf}) then yields the implicit equation for the anomalous dimensions
\be
  0 = \etax_\kappa  +\tI_\kappa^\xx(0) 
\label{etapert}
\ee
where  both integrals $\tI_\kappa^\nu(0)$ and   $\tI_\kappa^\td(0)$ depend linearly on $\eta_\kappa^\nu$ and $\eta_\kappa^\td$. We hence define
\begin{subequations}
  \begin{align}
   \tI_\kappa^\td(0) &= \tI^{\td\td}_\kappa \etad_\kappa + \tI^{\td\nu}_\kappa \etan_\kappa + \tI^{\td\xn}_\kappa  , \\
   \tI_\kappa^\nu(0) &= \tI^{\nu \td}_\kappa \etad_\kappa + \tI^{\nu \nu}_\kappa \etan_\kappa + \tI^{\nu\xn}_\kappa.
  \end{align}
\end{subequations}
 The explicit form of the various integrals is given by  \cite{Kloss12}:
\begin{subequations}
\begin{align}
\tI^{\td\td}_\kappa &= - \tg_\kappa \frac{ v_d}{2} \int_{0}^{\infty} \! \! d \tq \, \tq^{d+3} \frac{ r(\tq^2) \, \hat k_\kappa(\tq)}{\tf^\lambda_\kappa(\tq) (\hat l_\kappa(\tq))^3} , \\
\tI^{\td\nu}_\kappa &= \tg_\kappa \frac{ 3 v_d}{4} \int_{0}^{\infty} \! \! d \tq \, \tq^{d+5} \frac{ r(\tq^2) \,(\hat k_\kappa(\tq))^2}{\tf^\lambda_\kappa(\tq) (\hat l_\kappa(\tq))^4} ,  \\
\tI^{\td\xn}_\kappa &= \tg_\kappa \frac{ v_d}{2} \int_{0}^{\infty} \! \! d \tq \, \frac{\tq^{d+5} (\partial_{\tq^2} r(\tq^2)) }{\tf^\lambda_\kappa(\tq) (\hat l_\kappa(\tq))^4} 
  \hat k_\kappa(\tq) \times\nonumber\\
&\qquad\qquad \Bigl[ 3 \tq^2 \hat  k_\kappa(\tq) - 2 \hat l_\kappa(\tq) \Bigr] ,  \\
\tI^{\nu\td}_\kappa &= \tg_\kappa \frac{v_{d} }{4d} \int_{0}^{\infty} \! \! d \tq \, \frac{\tq^{d+1} r(\tq^2)}{\tf^\lambda_\kappa(\tq) (\hat l_\kappa(\tq))^3} \times \nonumber \\
& \Bigl[\tf^\lambda_\kappa(\tq) \tq \partial_{\tq} \hat l_\kappa(\tq) - \hat l_\kappa(\tq) \tq \partial_{\tq} \tf^\lambda_\kappa(\tq) -2 \tf^\lambda_\kappa(\tq)\hat l_\kappa(\tq)  \Bigr] ,  \\
\tI^{\nu\nu}_\kappa &= -\tg_\kappa \frac{v_d}{4 d} \!\! \int_{0}^{\infty} \!\! \! \! d \tq \frac{\tq^{d+3} r(\tq^2)}{\tf^\lambda_\kappa(\tq) (\hat l_\kappa(\tq))^3} \Bigl[\tf^\lambda_\kappa(\tq) \tq \partial_{\tq} \hat k_\kappa(\tq) \!\!\nonumber \\
& \qquad - \Bigl(2\tq \partial_{\tq} \tf^\lambda_\kappa(\tq) + (2-d) \tf^\lambda_\kappa(\tq)  \Bigr) \hat k_\kappa(\tq) \Bigr] , \\
\tI^{\nu\xn}_\kappa &= \! -\tg_\kappa \frac{v_d}{2 d}\!\! \int_{0}^{\infty} \!\! \! d \tq \, \frac{\tq^{d+3} \partial_{\tq^2} r(\tq^2)}{\tf^\lambda_\kappa(\tq) (\hat l_\kappa(\tq))^3} \times \nonumber\\
&\qquad \Bigl[\tq \partial_{\tq} \tf^\lambda_\kappa(\tq) \Bigl(-2 \tq^2 \hat k_\kappa(\tq)  + \hat l_\kappa(\tq)\Bigr) +\! \tf^\lambda_\kappa(\tq) \times\nonumber \\
& \Bigl( \tq^3 \partial_{\tq} \hat k_\kappa(\tq) - \tq \partial_{\tq} \hat l_\kappa(\tq)  +(d-2)\tq^2  \hat k_\kappa(\tq)  + 2  \hat l_\kappa(\tq)  \Bigr) \Bigr] ,
\end{align}
\label{integral}
\end{subequations}
where
\begin{subequations}
\begin{align}
\hat k_\kappa( \hat q) &= \hat f^\td_\kappa( \hat q)+\hat{w} \tq^{-2\rho}+ r( \hat q^2) , \\
\hat l_\kappa( \hat q) &= \hat q^2 ( \hat f^\nu_\kappa(\hat q)+ r( \hat q^2 )).
\end{align}
\end{subequations}
[Note two misprints in Eqs. (A4d) and (A4f) of Ref. \cite{Kloss12} corrected here].

To investigate the properties of the EWLR fixed points, we consider the limit $\hat{w} \rightarrow \infty$ and 
$\hat{g} \rightarrow 0$, at $\hat{g}\hat{w}^2$ fixed, that is $\hat{x}\to 1$ at fixed $\hat{y}$. As long as $\hat f^\nu_\kappa$, $ \hat f^D_\kappa$
and $\hat f^\lambda_\kappa$ are of order one in this limit (which holds by definition at LPA' and is verified below at NLO), it amounts to replacing in the various Eqs.  (\ref{integral}) $\hat k_\kappa( \hat q)$
by $\hat{w} \tq^{-2\rho}$. Accordingly, one observes that 
\begin{subequations}
\begin{align}
\tI^{\td\td}_\kappa &=\mathcal{O}(\hat{y}/\hat{w}),  \\
\tI^{\td\nu}_\kappa &= \mathcal{O}(\hat{y}), \\
\tI^{\td\xn}_\kappa &= \mathcal{O}(\hat{y}), \\
\tI^{\nu\td}_\kappa &=\mathcal{O}(\hat{y}/\hat{w}^2), \\
\tI^{\nu\nu}_\kappa &= \mathcal{O}(\hat{y}/\hat{w}), \\
\tI^{\nu\xn}_\kappa &=\mathcal{O}(\hat{y}/\hat{w}).
\end{align}
\end{subequations}
Let us check the behavior of the three functions $f_\kappa^\xx$ in this limit in the NLO approximation. The  NLO flow equations for the functions $\hat f^\nu_\kappa$, $ \hat f^D_\kappa$
and $\hat f^\lambda_\kappa$ (see \cite{Kloss12}) are of order
\begin{subequations}
\begin{align}
\partial_s \hat f^\td_\kappa( \hat q) &=\mathcal{O}(\hat{y}),  \\
\partial_s \hat f^\nu_\kappa(\hat q) &=\mathcal{O}(\hat{y}/\hat{w}), \\
\partial_s \tf^\lambda_\kappa(\tq) &= \mathcal{O}(\hat{y}/\hat{w}).
\end{align}
\end{subequations}
As a consequence, in the limit $\hat{x}\to 1$ at fixed $\hat{y}$, one has indeed
\begin{subequations}
\begin{align}
\hat f^\td_\kappa( \hat q) &=\mathcal{O}(1),  \\
 \hat f^\nu_\kappa(\hat q) & \to 1, \\
\tf^\lambda_\kappa(\tq) & \to 1.
\end{align}
\end{subequations}
Therefore, even if $\hat f^D_\kappa$ does not remain at bare level along the flow, the NLO and LPA' expressions for the anomalous dimensions Eqs. (\ref{etapert}) in this limit are identical and become
\begin{equation}
\etad_\kappa = \rho \hat{y}_\kappa \tilde{\eta}^{\td}(\rho, d) , \qquad
\etan_\kappa = \rho \hat{y}_\kappa (1-\hat{x}_\kappa) \tilde{\eta}^{\nu}(\rho, d) ,
\label{eq:etaslimit}
\end{equation}
with
\begin{subequations}
\begin{align}
\tilde{\eta}^{\td}(\rho, d) &=  -  6 v_d \int_{0}^{\infty} \! \! d \tq \,   
  \frac{ \tq^{d-1-4 \rho} (\partial_{\tq^2} r(\tq^2))  }{ [1 + r(\tq^2)]^4} , \\
\tilde{\eta}^{\nu}(\rho, d) &= \frac{2 v_d}{d}  \int_{0}^{\infty} \! \! d \tq \, 
  \frac{  \tq^{d-1-2 \rho} (\partial_{\tq^2} r(\tq^2))   \,(d-2-2 \rho)}{(1 + r(\tq^2))^3} .
\end{align}
\label{eq:etasInt}
\end{subequations}
Moreover the perturbative expansion of the NPRG flow equations for the couplings Eqs. (\ref{eq:minimalRG}) in the variables $\hat x_\kappa$ and $\hat y_\kappa$  become 
\begin{subequations}
\begin{align}
\partial_s \hat{x}_\kappa &= \rho  (\hat{x}_\kappa-1) (2 - \tilde{\eta}^{\td}(\rho, d) \hat{y}_\kappa ) +\mathcal{O}(\hat{x}_\kappa-1)^2 , \\
\partial_s \hat{y}_\kappa &= \hat{y}_\kappa ( d -2 + \rho (\tilde{\eta}^{\td}(\rho, d)\hat{y}_\kappa -4))+\mathcal{O}(\hat{x}_\kappa-1).
\end{align}
\label{eq:rescalLimit}
\end{subequations}

To study the stability of the EWLR fixed points, these flow equations can be linearized in the vicinity of a fixed point $(\hat{x}_*,\hat{y}_*)$ 
with $\hat{x}_*=1$ and the corresponding stability matrix reads
\begin{align}
\Omega &= 
\biggl(
\begin{array}{cc}
\frac{\partial (\partial_s \hat{x}_\kappa)}{\partial \hat{x}_\kappa} & \frac{\partial (\partial_s \hat{x}_\kappa)}{\partial \hat{y}_\kappa} \nonumber \\
\frac{\partial (\partial_s \hat{y}_\kappa)}{\partial \hat{x}_\kappa} & \frac{\partial (\partial_s \hat{y}_\kappa)}{\partial \hat{y}_\kappa} 
\end{array}
\biggr)\biggr|_{\hat{x}_\kappa=\hat{x}_*=1,\hat{y}_\kappa=\hat{y}_*}\\
&=\biggl(
\begin{array}{cc}
\rho(2-\hat{y}_* \tilde{\eta}^{\td}) & 0 \\
\frac{\partial (\partial_s \hat{y}_\kappa)}{\partial \hat{x}_\kappa}|_{\hat x_\kappa=1,\hat{y}_\kappa=\hat{y}_*} & d-2-4\rho+ 2 \rho \hat{y}_* \tilde{\eta}^{\td}
\end{array}
\biggr) .
\label{eq:linFlowMatrix}
\end{align}
To determine the component $\partial (\partial_s \hat y_\kappa)/\partial \hat x_\kappa$ requires to push the expansion (\ref{eq:rescalLimit}) of $\partial_s \hat y_\kappa$ to order $(\hat{x}_\kappa-1)$, but this
is not necessary for the study of the stability of fixed points with $\hat x_*=1$.
The expression (\ref{eq:linFlowMatrix}) of the stability matrix implies that for any fixed point with $\hat{x}_*=1$, the two 
eigenvalues,  which identify with the correction-to-scaling exponents, are $\omega_1=\rho(2-\hat{y}_* \tilde{\eta}^{\td})$ and $\omega_2=d-2-4\rho+ 2 \rho \hat{y}_* \tilde{\eta}^{\td}$.
We now discuss the two fixed point with $\hat{x}_*=1$.

\subsection*{EWLR1 fixed point}
The EWLR1 fixed point is located at $\hat{x}_* = 1$ and $\hat{y}_* = 0$.
The corresponding  correction-to-scaling exponents are
\begin{equation}
\omega_1 = 2 \rho ,  \qquad
\omega_2 =  \epsilon - 4 \rho,
\label{eq:ewlr1EW}
\end{equation}
 that are identical to the correction-to-scaling exponents obtained by JFT \cite{Frey99,Janssen99}. Consequently, we recover the same stability conditions, namely  the EWLR1 fixed point is always attractive in the $\hat{x}$-direction, whereas it is attractive in the $\hat{y}$-direction for $d > 2 (1+2 \rho)$ and repulsive for $d < 2 (1+2 \rho)$.

\subsection*{EWLR2 fixed point}
The coordinates of the EWLR2 fixed point are
\begin{equation}
\hat{x}_* = 1, \qquad \hat{y}_* = \frac{4}{\tilde{\eta}^{\td}(\rho, d)} \left( 1 - \frac{\epsilon}{4 \rho}\right) ,
\label{eq:ewlr2pos}
\end{equation}
 and the corresponding correction-to-scaling exponents are given by
\begin{equation}
\omega_1 =  \epsilon - 2 \rho ,  \qquad
\omega_2 =  4 \rho - \epsilon ,
\label{eq:ewlr2EW}
\end{equation}
which  again identify with the correction-to-scaling  exponents found perturbatively \cite{Frey99,Janssen99}. Checking  that $\tilde{\eta}^{\td}(\rho, d)$ is always positive, we deduce the same stability conditions as JFT.
 The EWLR2 fixed point is always attractive in the $\hat{x}$-direction, and
  EWLR1 and EWLR2 exchange their stability in the $\hat{y}$-direction when EWLR2 crosses EWLR1 at $ d_{\ewlr} = 2 (1 + 2 \rho)$.

\renewcommand{\theequation}{B\arabic{equation}}
\section*{Appendix B: Cutoff dependence}
\setcounter{equation}{0}

In this Appendix, we discuss the dependence of our results in the regulator function, which can be tested
 {\it via} the variation of the (positive) parameter $\alpha$ in Eq. (\ref{eq:expReg}). 
Of course,  physical observables computed from the exact NPRG equation (\ref{eq:dkgam}) do not depend on the choice of regulator.
However, any approximation induces a spurious dependence in this regulator, which can be used to 
 test the quality of the approximation.

The $\alpha$ dependence of the critical exponent $\chi_{\sr}$ of the SR fixed point at the NLO approximation has been studied in details in Ref. \cite{Kloss12}. The exponent $\chi_{\sr}$ is exact in $d=1$ (no $\alpha$-dependence), depends very weakly on $\alpha$ in $d=2$,
 with an optimal value $\chi_{\sr}\simeq 0.373$, and somewhat more in $d=3$, with an optimal value $\chi_{\sr}\simeq 0.179$. In this work, we used $\alpha=4$ in $d=2,3$ which belongs to the interval in $\alpha$ where the variations of the critical exponents 
with this parameter are minimal, and their values very close to the optimal ones, see \cite{Kloss12}.
Increasing further the dimension above $d \gtrsim 3.5$, an increased cutoff
dependence was observed, with even unphysical negative  $\chi_{\sr}$  values for small
$\alpha$ in $d=4$.
 This clearly signals that the NLO approximation becomes less accurate
 as the dimension grows, and quantitatively unreliable for $d \gtrsim 3.5$.
In particular, we here chose $\alpha=10$ in $d=4$ in order to get a positive exponent
 $\chi_{\sr}$, which is of the same order as the optimal value obtained at LO, see \cite{Kloss12}.

Let us now review the sensitivity of the results in the LR sector presented in this work
 with respect to a variation of the cutoff function, that is of $\alpha$.
First, the boundary line $\rho_{\sr}(d)$ separating the LR and SR phases is entirely determined
 by the value of $\chi_{\sr}$, so its dependence on $\alpha$ can be directly inferred from 
 that of $\chi_{\sr}$ discussed above.
Then,  the critical exponents of the LR dominated phases
 (EWLR1, EWLR2 and LR)  are exact and thus independent of $\alpha$.
The same holds true for the correction-to-scaling exponents Eqs.\ (\ref{eq:ewlr1EW},\ref{eq:ewlr2EW})
 at the EWLR fixed points, and thus for their stabilities.
Indeed, only  the coordinate $\hat{y}_*$ of EWLR2 depends on $\alpha$ through
$\tilde{\eta}^{\td}$ (Eq.\ (\ref{eq:etasInt})) and we checked that $\tilde{\eta}^{\td}$
 remains positive for all values of $\alpha$.

Finally, the coordinates of the LR, SR and T fixed points do depend on $\alpha$.
However, below $d \simeq 3.5$, and also in $d=4$ provided a sufficiently large value of $\alpha$
 is chosen (to ensure that $\chi_{\sr}>0$), the qualitative stucture of the flow diagram with
 the relative positions of T, SR and LR presented in FIG. \ref{fig:2d-flow}, \ref{fig:3d-flow} and
  \ref{fig:4d-flow} is preserved, together with the stability change of SR and LR at
$\chi_{\sr}=\chi_{\lr}$.

\end{appendix}

\bibliography{biblioKPZ}
\bibliographystyle{apsrev4-1}

\end{document}